\documentclass[11pt]{article}
\usepackage[top=2cm,left=1.5cm,right=1.5cm]{geometry}
\usepackage[numbers,square]{natbib}

\usepackage[english]{babel}
\usepackage{epsfig}
\usepackage{amssymb}
\usepackage{amsfonts}
\usepackage{rotating}
\usepackage{amsmath}
\usepackage{fancyhdr}
\usepackage{lineno}
\usepackage{subfigure}
\usepackage{graphics}
\usepackage{geometry}
\usepackage{pstricks}
\usepackage{color}

\let\a   = \alpha     \let\b = \beta    \let\g = \gamma    \let\d = \delta
       \let\h = \eta

\let\s   = \sigma            
            
\newcommand{\beq}{\begin{equation}}
\newcommand{\eeq}{\end{equation}}

\newcommand{\beqa}{\begin{eqnarray}}
\newcommand{\eeqa}{\end{eqnarray}}

\newcommand{\bea}{\begin{eqnarray}}
\newcommand{\eea}{\end{eqnarray}}

\newcommand{\bann}{\begin{eqnarray*}}
\newcommand{\eann}{\end{eqnarray*}}

\def\nn{\nonumber}

\newcommand{\bmi}{\begin{minipage}}
\newcommand{\emi}{\end{minipage}}

\newcommand{\pd}{\partial}
\newcommand{\ksl}{k \! \! \!  /}

\begin{document}

\begin{center}
\vspace{4.cm}

{\bf \large  Dilaton Interactions and the Anomalous Breaking \\ of Scale Invariance of the Standard Model } 

\vspace{1cm}
{\bf Claudio Corian\`{o}, Luigi Delle Rose, Antonio Quintavalle and Mirko Serino}

\vspace{1cm}

{\it Dipartimento di Matematica e Fisica "Ennio De Giorgi"
\\Universit\`{a} del Salento 
\\ and \\ INFN-Lecce \\ Via Arnesano 73100, Lecce, Italy\footnote{claudio.coriano@unisalento.it, luigi.dellerose@le.infn.it, 
antonio.quintavalle@le.infn.it, mirko.serino@le.infn.it}
}\\
\vspace{.5cm}
\begin{abstract}

We discuss the main features of dilaton interactions for fundamental and effective dilaton fields. In particular, we elaborate on the 
various ways in which dilatons can couple to the Standard Model and on the role played by the conformal anomaly as a way to 
characterize their interactions.
In the case of a dilaton derived from a metric compactification (graviscalar), we present the structure of the radiative corrections 
to its decay into two photons, a photon and a $Z$, two $Z$ gauge bosons and two gluons, together with their renormalization 
properties. We prove that, in the electroweak sector, the renormalization of the theory is guaranteed only if the Higgs is 
conformally coupled. For such a dilaton, its coupling to the trace anomaly is quite general, and determines, for instance, 
an enhancement of its decay rates into two photons and two gluons.   
We then turn our attention to theories containing a non-gravitational (effective) dilaton, which, in our perturbative analysis, 
manifests as a pseudo-Nambu Goldstone mode of the dilatation current ($J_D$). The infrared coupling of such a state to the 
two-photons and to the two-gluons sector, and the corresponding anomaly enhancements of its decay rates in these channels, 
is critically analyzed.

\end{abstract}
\end{center}

\newpage

\section{Introduction} 

Dilatons are part of the low energy effective action of several different types of theories, from string theory to theories with compactified extra dimensions, but they may appear also in appropriate bottom-up constructions. For instance, in scale-invariant extensions of the Standard Model, the introduction of a dilaton field allows to recover scale invariance, which is violated by the Higgs potential, by introducing a new, enlarged, Lagrangian. This is characterized both by a spontaneous breaking of the conformal and of the electroweak symmetries.  

In this case, one can formulate simple scale-invariant extensions of the potential which can accomodate, via spontaneous breakings, 
two separate scales: the electroweak scale ($v$), related to the vev of the Higgs field, and the conformal breaking scale 
($\Lambda$), related to the vev of a new field $\Sigma=\Lambda + \rho$, with $\rho$ being the dilaton. The second scale can be 
fine-tuned in order to proceed with a direct phenomenological analysis and is, therefore, of outmost relevance in the search for new 
physics at the LHC.

In a bottom-up approach, and this will be one of the main points that we will address in our analysis, the dilaton of the effective 
scale-invariant Lagrangian can also be interpreted as a composite scalar, with the dilatation current taking the role of an operator 
which interpolates between this state and the vacuum. We will relate this intepretation to the appearance of an anomaly pole in the 
correlation function involving the dilatation current ($J_D$)
and two neutral currents ($V, V'$) of the Standard Model, providing evidence, in the ordinary perturbative picture, in favour of such 
a statement.   

One of the main issues which sets a difference between the various types of dilatons is, indeed, the contribution coming from the 
anomaly, which is expected to be quite large. Dilatons obtained from compactifications with large extra dimensions and a low gravity 
scale, for instance, carry this coupling, which is phenomenologically relevant. 
The same coupling is present in the case of an effective dilaton, appearing as a Goldstone mode of the dilatation current, with some 
differences that we will specify in a second part of our work. The analysis will be carried out in analogy to the pion case, which in 
a perturbative picture is associated with the appearance of an anomaly pole in the $AVV$ diagram (with $A$ being the axial current).

Our work is organized as follows. In a first part we will characterize the leading one-loop interactions of a dilaton derived from a 
Kaluza-Klein compactification of the gravitational metric. The setup is analogous to that presented in 
\cite{Han:1998sg,Giudice:2000av} for a compactified theory with large extra dimensions and it involves all the neutral currents of 
the Standard  Model. We present also a discussion of the same interaction in the QCD case for off-shell gluons.    

These interactions are obtained by tracing the $TVV'$ vertex, with $T$ denoting the (symmetric and conserved) 
energy-momentum tensor (EMT) of the Standard Model. This study is accompanied by an explicit proof of the renormalizability of these 
interactions in the case of a conformally coupled Higgs scalar.

In a second part then we turn our discussion towards models in which dilatons are introduced from the ground up, starting with simple 
examples which should clarify - at least up to operators of dimension 4 - how one can proceed with the formulation of scale-invariant 
extensions of the Standard Model.  Some of the more technical material concerning this point has been left to the appendices, where 
we illustrate the nature of the coupling of the dilaton to the mass dependent terms of the corresponding Lagrangian. The goal of 
these technical additions is to clarify that a fundamental (i.e. not a composite) dilaton, in a {\em classical} scale invariant 
extension of a given Lagrangian, does not necessarily couple to the anomaly, but only to massive states, exactly as in the Higgs 
case. For an effective dilaton, instead,  the Lagrangian is derived at tree level on the basis of classical scale invariance, as for 
a fundamental dilaton, which needs to be modified with the addition of an anomalous contribution, due to the composite nature of the 
scalar, in close analogy to the pion case.   

As we are going to show, if the dilaton is a composite state, identified with the anomaly pole of 
the $J_D VV$ correlator, an infrared coupling of this pole (i.e. a nonzero residue) is necessary in order to claim the presence of an 
anomaly enhancement in the $VV$ decay channel, with the $VV$ denoting on-shell physical asymptotic states, in a typical $S$-matrix 
approach. Here our reasoning follows quite closely the chiral anomaly case, where the anomaly pole of the $AVV$ diagram, which 
describes the pion exchange between the axial vector ($A$) and the vector currents, is infrared coupled only if $V$ denote physical 
asymptotic states. 
 
Clearly, our argument relies on a perturbative picture and is, in this respect, admittedly limited, forcing this issue to be resolved 
at experimental level, as in the pion case. We recall that in the pion the enhancement is present in the di-photon channel and not in 
the 2-gluon decay channel. 
 
Perturbation theory, in any case, allows to link the enhancement of a certain dilaton production/decay channel, to the virtuality of 
the gauge currents in the initial or the final state. 
  
We conclude with a discussion of the possible phenomenological implications of our results at the level of anomaly-enhanced dilaton 
decays, after pointing out the difference between the various ways in which the requirement of scale invariance (classical or 
quantum) can be realized in a typical scale-invariant extension of the Standard Model Lagrangian.

\subsection{The energy momentum tensor}

We start with a brief summary of the structure of the Standard Model interactions with a $4D$ gravitational background, which is 
convenient in order to describe both the coupling of the graviscalar dilaton, emerging from the Kaluza-Klein compactification, and of 
a graviton at tree level and at higher orders.   
In the background metric $g_{\mu\nu}$ the action takes the form 
\beq S = S_G + S_{SM} + S_{I}= -\frac{1}{\kappa^2}\int d^4 x \sqrt{-g}\, R + \int d^4 x
\sqrt{-g}\mathcal{L}_{SM} + \chi \int d^4 x \sqrt{-g}\, R \, H^\dag H      \, ,
\eeq
where $\kappa^2=16 \pi G_N$, with $G_N$ being the four dimensional Newton's constant and $\mathcal H$ is the Higgs doublet.
We recall that the EMT in our conventions is defined as
\beq  T_{\mu\nu}(x)  = \frac{2}{\sqrt{-g(x)}}\frac{\d [S_{SM} + S_I ]}{\d g^{\mu\nu}(x)},
\eeq
or, in terms of the SM Lagrangian, as
\beq \label{TEI spaziocurvo}
\frac{1}{2} \sqrt{-g} T_{\mu\nu}{\equiv} \frac{\pd(\sqrt{-g}\mathcal{L})}
{\pd g^{\mu\nu}} - \frac{\pd}{\pd x^\s}\frac{\pd(\sqrt{-g}\mathcal{L})}{\pd(\pd_\s g^{\mu\nu})}\, ,
\eeq
which is classically covariantly conserved   ($g^{\mu\rho}T_{\mu\nu; \rho}=0$).  In flat spacetime, the covariant derivative is 
replaced by the ordinary derivative, giving the ordinary conservation equation ($ \pd_\mu T^{\mu\nu} = 0$).

We use the convention $\eta_{\mu\nu}=(1,-1,-1,-1)$ for the metric in flat spacetime, parameterizing its deviations 
from the flat case as
\beq\label{QMM} g_{\mu\nu}(x) \equiv \h_{\mu\nu} + \kappa \, h_{\mu\nu}(x)\,,\eeq
with the symmetric rank-2 tensor $h_{\mu\nu}(x)$ accounting for its fluctuations.

 In this limit, the coupling of the Lagrangian to gravity is given by the term
\beq\label{Lgrav} \mathcal{L}_{grav}(x) = -\frac{\kappa}{2}T^{\mu\nu}(x)h_{\mu\nu}(x). \eeq
In the case of theories with extra spacetime dimensions the structure of the corresponding Lagrangian can be found in 
\cite{Han:1998sg,Giudice:2000av}. For instance, in the case of a compactification over a $S_1$ circle of a 5-dimensional theory to 
4D, equation (\ref{Lgrav}) is modified in the form 
\beq
\label{Lgrav1} \mathcal{L}_{grav}(x) = -\frac{\kappa}{2} T^{\mu\nu}(x) \left(h_{\mu\nu}(x) + \rho(x) \,  \eta_{\mu\nu} \right) 
\eeq
which is sufficient in order to describe dilaton $(\rho)$ interactions with the fields of the Standard Model at leading order in 
$\kappa$, as in our case. In this case the graviscalar field $\rho$ is related to the $g_{55}$ component of the 5D metric and 
describes its massless Kaluza-Klein mode. The compactification generates an off-shell coupling of $\rho$ to the trace of the 
symmetric EMT. Notice that in this construction the fermions are 
assumed to live on the 4D brane and their interactions can be described by the ordinary embedding of the fermionic Lagrangian of the 
Standard Model to a curved 4D gravitational background.  We use the spin connection $\Omega$ induced by the
curved metric $g_{\mu\nu}$. This allows to define a spinor derivative $\mathcal{D}$ which transforms covariantly under local Lorentz 
transformations. If we denote with $\underline{a},\underline{b}$ the Lorentz indices of a local free-falling frame, and denote with
$\s^{\underline{a}\underline{b}}$ the generators of the Lorentz group in the spinorial representation, the spin connection takes 
the form
\beq
 \Omega_\mu(x) = \frac{1}{2}\s^{\underline{a}\underline{b}}V_{\underline{a}}^{\,\nu}(x)V_{\underline{b}\nu;\mu}(x)\, ,
\eeq
where we have introduced the vielbein $V_{\underline{a}}^\mu(x)$. The covariant derivative of a spinor in a given representation
$(R)$ of the gauge symmetry group, expressed in curved $(\mathcal{D}_{\mu})$ coordinates is then given by
\beq \mathcal{D}_{\mu} = \frac{\pd}{\pd x^\mu} + \Omega_\mu   + A_\mu,\eeq
where $A_\mu\equiv A_\mu^a\, T^{a  (R)}$ are the gauge fields and $T^{a (R)}$ the group generators,
giving a Lagrangian of the form
\beqa \mathcal{L}& = & \sqrt{-g} \bigg\{\frac{i}{2}\bigg[\bar\psi\g^\mu(\mathcal{D}_\mu\psi)
 - (\mathcal{D}_\mu\bar\psi)\g^\mu\psi \bigg] - m\bar\psi\psi\bigg\}\, .       
\eeqa
The derivation of the complete dilaton/gauge/gauge vertex in the Standard Model requires the computation of the trace of the EMT 
${T^\mu}_\mu$ (for the tree-level contributions), and of a large set of 1-loop 3-point functions. 
These are diagrams characterized by the insertion of the trace into 2-point functions of gauge currents. 
The full EMT is given by a minimal tensor $T_{Min}^{\mu\nu}$ (without improvement) and by a term of improvement, 
$T_I^{\mu\nu}$, originating from $S_I$, as
\beqa
T^{\mu\nu} = T_{Min}^{\mu\nu} + T_I^{\mu\nu} \,,
\eeqa
where the minimal tensor is decomposed into gauge, ghost, Higgs, Yukawa and gauge fixing (g.fix.) contributions which can be found in 
\cite{Coriano:2011zk} 
\beqa
T_{Min}^{\mu\nu} = 
T_{gauge}^{\mu\nu} + T_{ferm.}^{\mu\nu} + T_{Higgs}^{\mu\nu} + T_{Yukawa}^{\mu\nu} + T_{g.fix.}^{\mu\nu} + T_{ghost}^{\mu\nu}.
\eeqa
Concerning the structure of the EMT of improvement, we introduce the ordinary parametrizations of the Higgs field  
\beq
\label{Higgsparam}
H = \left(\begin{array}{c} -i \phi^{+} \\ \frac{1}{\sqrt{2}}(v + h + i \phi) \end{array}\right)
\eeq
and of its conjugate $H^\dagger$, expressed in terms of $h$, $\phi$ and $\phi^{\pm}$, corresponding to the physical Higgs and the 
Goldstone bosons of the $Z$ and $W^{\pm}$ respectively. As usual, $v$ denotes the Higgs vacuum expectation value. This expansion 
generates a non-vanishing EMT, induced by $S_I$, given by
\bea
\label{Timpr}
T^{\mu\nu}_I = - 2 \chi (\partial^\mu \partial^\nu - \eta^{\mu \nu} \Box) H^\dag H = 
- 2 \chi (\partial^\mu \partial^\nu - \eta^{\mu\nu} \Box) \left( \frac{h^2}{2} + \frac{\phi^2}{2} + \phi^+ \phi^- + v \, h\right)\, .
\eea
Notice that this term is generated by a Lagrangian which does not survive the flat spacetime limit. We are going to show by an 
explicit computation that $T_I$, if properly included with $\chi=1/6$, guarantees the renormalizability of the model.

\section{One loop electroweak corrections to dilaton-gauge-gauge vertices} 

In this section we will present results for the structure of the radiative corrections to the dilaton/gauge/gauge vertices in the 
case of two photons, photon/$Z$ and $Z Z$ gauge currents. We have included in Appendix \ref{rules} the list of the relevant tree 
level interactions extracted from the SM Lagrangian  
introduced above and which have been used in the computation of these corrections. We identify three classes of contributions, 
denoted as $\mathcal{A}$, $\Sigma$ and $\Delta$, with the $\mathcal A$-term coming from the conformal anomaly while the $\Sigma$ and 
$\Delta$ terms are related to the exchange of fermions, gauge bosons and scalars (Higgs/Goldstones). The separation between the 
anomaly part and the remaining terms is typical of the $TVV'$ interaction. In particular one can check that in a 
mass-independent renormalization scheme, such as Dimensional Regularization with minimal subtraction, this separation can be verified 
at least at one loop level and provides a realization of the (anomalous) conformal Ward identity 
\beq
\Gamma^{\alpha\beta}(z,x,y) 
\equiv \eta_{\mu\nu} \left\langle T^{\mu\nu}(z) V^{\alpha}(x) V^{\beta}(y) \right\rangle 
= \frac{\delta^2 \mathcal A(z)}{\delta A_{\alpha}(x) \delta A_{\beta}(y)} + \left\langle {T^\mu}_\mu(z) V^{\alpha}(x) V^{\beta}(y) 
\right\rangle,
\label{traceid1}
\eeq
where we have denoted by $\mathcal A(z)$ the anomaly and $A_{\alpha}$ the gauge sources coupled to the current $V^{\alpha}$. Notice 
that in the expression above $\Gamma^{\alpha\beta}$ denotes a generic 
dilaton/gauge/gauge vertex, which is obtained form the $TVV'$ vertex by tracing the spacetime indices $\mu\nu$. A simple way to test 
the validity of (\ref{traceid1}) is to compute the renormalized vertex $\langle T^{\mu\nu} V^\alpha V'^\beta\rangle$ (i.e. the 
graviton/gauge/gauge vertex) and perform afterwards its 4-dimensional trace. This allows to identify the left-hand-side of this 
equation. On the other hand, the insertion of the trace of $T^{\mu\nu}$ (i.e. $T^\mu_\mu$ )into a two point function $VV'$, allows to 
identify the second term on the right-hand-side of (\ref{traceid1}),  $\langle T^{\mu}_\mu(z) V^{\alpha}(x) V^{\beta}(y)\rangle$. 
The difference between the two terms so computed can be checked to correspond to the $\mathcal A$-term, obtained by two 
differentiations of the  anomaly functional $\mathcal A$. 
We recall that, in general, when scalars are conformally coupled, this takes the form
\beqa \label{TraceAnomaly}
\mathcal A(z)
&=& \sum_{i} \frac{\beta_i}{2 g_i} \, F^{\alpha\beta}_i(z) F^i_{\alpha\beta}(z) +... \,,
\eeqa
where $\beta_i$ are clearly the mass-independent $\beta$ functions of the gauge fields
and $g_i$ the corresponding coupling constants, while the ellipsis refer to curvature-dependent terms. 
We present explicit results starting for the $\rho VV'$ vertices ($V,V' = \gamma, Z$), denoted as $\Gamma_{VV'}^{\alpha \beta}$, 
which are decomposed in momentum space in the form 
\beq
\Gamma_{VV'}^{\alpha \beta}(k,p,q) = (2\, \pi)^4\, \delta(k-p-q) \frac{i}{\Lambda} 
\left( \mathcal A^{\alpha \beta}(p,q) + \Sigma^{\alpha \beta}(p,q) + \Delta^{\alpha \beta}(p,q)\right),
\eeq
where 
\beq
\mathcal A^{\alpha \beta}(p,q) = \int d^4 x\, d^4 y \, e^{i p \cdot x + i q\cdot y}\, 
\frac{\delta^2 \mathcal A(0)}{\delta A^\alpha(x)\delta A^\beta(y)}
\eeq
and 
\beq
 \Sigma^{\alpha \beta}(p,q) +  \Delta^{\alpha \beta}(p,q) = \int d^4 x\, d^4 y\, e^{ i p \cdot x + i q\cdot y}\, 
\left\langle {T^\mu}_\mu(0) V^\alpha(x) V^\beta(y) \right\rangle \,.
\eeq
We have denoted with $\Sigma^{\alpha \beta}$ the cut vertex contribution to $\Gamma^{\alpha\beta}_{\rho VV'}$, 
while $\Delta^{\alpha \beta}$ includes the dilaton-Higgs mixing on the dilaton line, as shown in Fig. \ref{figuremix}.
Notice that $\Sigma^{\alpha \beta}$ and $\Delta^{\alpha \beta}$ take contributions in two cases, specifically if the theory has an 
explicit (mass dependent) breaking and/or if the scalar - which in this case is the Higgs field - is not conformally coupled. 
The $\mathcal A^{\alpha\beta}(p,q)$ represents the conformal anomaly while $\Lambda$ is dilaton interaction scale.

\subsection{The $\rho\gamma\gamma$ vertex}

The interaction between a dilaton and two photons is identified by the diagrams in Figs. \ref{figuretriangle},\ref{figuretadpole},\ref{figuremix} and is summarized by the expression 
\beqa
\Gamma_{\gamma \gamma}^{\alpha\beta}(p,q) = \frac{i}{\Lambda} \bigg[ \mathcal A^{\alpha\beta}(p,q) + 
\Sigma^{\alpha\beta}(p,q) + \Delta^{\alpha\beta}(p,q) \bigg] \,,\eeqa
with the anomaly contribution given by
\beqa
\label{Agammagamma}
\mathcal A^{\alpha\beta} =   \frac{\alpha}{\pi}\, \bigg[ -\frac{2}{3}\sum_{f} Q_{f}^2 + \frac{5}{2}  + 6\,\chi\bigg]\,
 u^{\alpha\beta}(p,q) \stackrel{\chi\rightarrow\frac{1}{6}}{=} - 2\, \frac{\beta_e}{e}\, u^{\alpha\beta}(p,q) \, ,
\eeqa
where
\bea
\label{utensor}
u^{\alpha\beta}(p,q) = (p\cdot q) \eta^{\alpha\beta} - q^{\alpha}p^{\beta} \,,
\eea
and the explicit scale-breaking term $\Sigma^{\alpha \beta}$ which splits into
\beqa
\label{Sigmagammagamma}
\Sigma^{\alpha\beta}(p,q) = \Sigma_F^{\alpha\beta}(p,q) +  \Sigma_B^{\alpha\beta}(p,q) +\Sigma_I^{\alpha\beta}(p,q) \,.
\eeqa
We obtain for the on-shell photon case ($p^2 = q^2 = 0$)
\beqa
\Sigma_F^{\alpha\beta}(p,q) 
&=& 
 \frac{\alpha}{\pi}\, \sum_f Q_f^2 m_f^2 \left[ \frac{4}{s} + 2 \left(\frac{4 
m_f^2}{s}-1\right) \mathcal C_0\left(s,0,0,m_f^2,m_f^2,m_f^2 \right)\right]\, u^{\alpha\beta}(p,q) \, , \nn \\
\Sigma_B^{\alpha\beta}(p,q) 
&=& 
 \frac{\alpha}{\pi}\, \left[ 6 M_W^2 \left(1-2\frac{M_W^2}{s}\right)  
\mathcal C_0(s,0,0,M_W^2,M_W^2,M_W^2) - 6 \frac{M_W^2}{s} - 1 \right]\, u^{\alpha\beta}(p,q) \, , \nn \\
\Sigma_I^{\alpha\beta}(p,q) 
&=& 
  \frac{\alpha}{\pi}\, 6 \chi \bigg[ 2 M_W^2 \mathcal C_0\left(s,0,0,M_W^2,M_W^2,M_W^2\right) \,
u^{\alpha\beta}(p,q) \nonumber \\
&-& 
M_W^2\, \frac{s}{2}\, \mathcal C_0(s,0,0,M_W^2,M_W^2,M_W^2) \, \eta^{\alpha \beta} \bigg]\, ,
\eeqa
while the mixing contributions are given by
\bea
\label{DeltaHgammagamma}
\Delta^{\alpha\beta}(p,q) 
&=&
\frac{\alpha}{\pi (s - M_H^2)} 6 \chi \bigg\{ 2 \sum_f Q_f^2 m_f^2  \bigg[ 2 + (4 m_f^2 -s ) \mathcal C_0(s,0,0,m_f^2,m_f^2,m_f^2) 
\bigg] \nn \\
&+& 
 M_H^2 + 6 M_W^2 + 2 M_W^2 (M_H^2 + 6 M_W^2 -4 s) \mathcal C_0(s,0,0,M_W^2,M_W^2,M_W^2)\bigg\} u^{\alpha \beta}(p,q) \nn \\
&+&  
\frac{\alpha}{\pi} 3 \chi s \, M_W^2 \, \mathcal C_0(s,0,0,M_W^2,M_W^2,M_W^2) \eta^{\alpha \beta} \, ,
\eea
with $\alpha$ the fine structure constant. The scalar integrals are defined in Appendix \ref{scalars}.
The $\Sigma$'s  and $\Delta$ terms are the contributions obtained from the insertion on the photon 2-point function of the trace of 
the EMT, ${T^\mu}_\mu$. Notice that $\Sigma_I$ includes all the trace insertions which originate from the terms of improvement $T_I$ 
except for those which are bilinear in the Higgs-dilaton fields and 
which have been collected in $\Delta$. The analysis of the Ward and Slavnov-Taylor identities for the graviton-vector-vector correlators shows that these can be consistently solved only if we include the graviton-Higgs mixing on the graviton line. 

We have included contributions proportional both to fermions ($F$) and boson ($B$) loops, beside the $\Sigma_I$.
A conformal limit on these contributions can be performed by sending to zero all the mass terms, which is equivalent to sending 
the vev $v$ to zero and requiring a conformal coupling of the Higgs $(\chi=1/6)$. 
In the $v\to 0$ limit, but for a generic parameter $\chi$, we obtain 
\beq
\lim_{v\to 0} \left( \Sigma^{\a\b} + \Delta^{\a\b} \right)
= \lim_{v\to 0} \left(\Sigma_{B}^{\a\b}  + \Sigma_{I}^{\a\b}\right) 
=  \frac{\alpha}{\pi} (6 \chi -1) u^{\alpha \beta}(p,q),
\eeq
which, in general, is non-vanishing.
Notice that, among the various contributions, only the exchange of a boson or the term of improvement contribute in this limit and 
their sum vanishes only if the Higgs is conformally coupled $(\chi = \frac{1}{6})$. \\

Finally, we give the decay rate of the dilaton into two on-shell photons in the simplified case in which we remove the term of 
improvement by sending $\chi \to 0$
\beqa
\Gamma(\rho \rightarrow \gamma\gamma) 
&=&
\frac{\alpha^2\,m_{\rho}^3}{256\,\Lambda^2\,\pi^3} \, \bigg| \beta_{2} + \beta_{Y} 
-\left[ 2 + 3\, x_W  +3\,x_W\,(2-x_W)\,f(x_W) \right]
+ \frac{8}{3} \, x_t\left[1 + (1-x_t)\,f(x_t) \right] \bigg|^2 \,,
\label{PhiGammaGamma} 
\eeqa
where the contributions to the decay, beside the anomaly term, come from the $W$ and the fermion (top) loops
and $\beta_2 (= 19/6)$ and $\beta_Y (= -41/6)$ are the $SU(2)$ and $U(1)$ $\beta$ functions respectively.
Here, as well as in the other decay rates evaluated all through the paper, the $x_i$ are defined as
\beq \label{x}
x_i = \frac{4\, m_i^2}{m^2_\rho} \, ,
\eeq
with the index "$i$" labelling the corresponding massive particle, and $x_t$ denoting the contribution from the top quark,
which is the only massive fermion running in the loop.
The function $f(x)$ is given by
\beqa
\label{fx}
f(x) = 
\begin{cases}
\arcsin^2(\frac{1}{\sqrt{x}})\, , \quad \mbox{if} \quad \,  x \geq 1 \\ 
-\frac{1}{4}\,\left[ \ln\frac{1+\sqrt{1-x}}{1-\sqrt{1-x}} - i\,\pi \right]^2\, , \quad \mbox{if} \quad \, x < 1.
\end{cases}
\eeqa
which originates from the scalar three-point master integral through the relation 
\beq \label{C03m}
C_0(s,0,0,m^2,m^2,m^2) = - \frac{2}{s} \, f(\frac{4\,m^2}{s}) \, .
\eeq

\subsection{The $\rho\gamma Z$ vertex} 
%
The interaction between a dilaton, a photon and a $Z$ boson is described by the $\Gamma^{\alpha \beta}_{\gamma Z}$ correlation 
function (Figs. \ref{figuretriangle},\ref{figuretadpole},\ref{figuremix}). In the on-shell case, with the kinematic defined by
\beq
p^2 = 0 \, \quad q^2 = M_Z^2 \, \quad k^2 = (p+q)^2 = s \,,
\eeq
the vertex $\Gamma^{\alpha \beta}_{\gamma Z}$ is expanded as 
\bea
\Gamma^{\alpha \beta}_{\gamma Z} &=& \frac{i}{\Lambda} \bigg[ \mathcal A^{\alpha\beta}(p,q)
+ \Sigma^{\alpha\beta}(p,q) + \Delta^{\alpha\beta}(p,q) \bigg] \nn \\
&=&
\frac{i}{\Lambda}\, \Bigg\{\,
\left[ \frac{1}{2}\,\left(s - M^2_Z\right)\,\eta^{\alpha\beta} - q^\alpha\,p^\beta\right] \, 
\left( \mathcal A_{\gamma Z} +  \Phi_{\gamma Z}(p,q)\right)
+\eta^{\alpha\beta}\, \Xi_{\gamma Z}(p,q) \Bigg\} \, .
\eea
The anomaly contribution is
\beq
\mathcal A_{\gamma Z} =  \frac{\alpha}{\pi\,s_w c_w}\,\left[-\frac{1}{3}\sum_{f}C^f_v\,Q_f + \frac{1}{12}\,(37-30s_w^2)
+ 3\, \chi\, (c_w^2 - s_w^2) \right]\, ,
\eeq
where $s_w$ and $c_w$ to denote the sine and cosine of the $\theta$-Weinberg angle.
Here $\Delta^{\alpha \beta}$ is the external leg correction on the dilaton line and the form factors $\Phi(p,q)$ and $\Xi(p,q)$ 
are introduced to simplify the computation of the decay rate and decomposed as
\bea
\Phi_{\gamma Z}(p,q)  &=& \Phi^{\Sigma}_{\gamma Z}(p,q) + \Phi^{\Delta}_{\gamma Z}(p,q) \,, \nn \\
\Xi_{\gamma Z}(p,q) &=& \Xi^{\Sigma}_{\gamma Z}(p,q) + \Xi^{\Delta}_{\gamma Z}(p,q) \,,
\eea
in order to distinguish the contributions to the external leg corrections ($\Delta$) from those to the cut vertex ($\Sigma$).
They  are given by
\beqa
\Phi^{\Sigma}_{\gamma Z}(p,q)
&=& 
\frac{\alpha}{\pi\, s_w\,c_w}\, \Bigg\{
\sum_f C_v^f \, Q_f \, \Bigg[\frac{2\, m_f^2}{s - M_Z^2} + \frac{2 m_f^2 \, M_Z^2}{(s - M_Z^2)^2}\, \mathcal D_0(s,M_Z^2,m_f^2,m_f^2)
\nonumber \\
&& \hspace{-20mm}
- m_f^2\, \left(1 - \frac{4\, m_f^2}{s - M_Z^2}\right)\, \mathcal C_0(s,0,M_Z^2,m_f^2,m_f^2,m_f^2) \Bigg]
-  \Bigg[ \frac{M_Z^2}{2\,\left(s - M_Z^2\right)}\, (12\, s_w^4 - 24\, s_w^2 + 11)
\nonumber\\
&& \hspace{-20mm}
\frac{M_Z^2}{2\,\left(s - M_Z^2 \right)^2} 
\left[2\, M_Z^2\,\Big(6\, s_w^4 - 11\, s_w^2 + 5 \Big)- 2\, s_w^2\, s + s \right]\, \mathcal D_0(s,M_Z^2,M_W^2,M_W^2)
\nonumber \\
&& \hspace{-20mm}
+ \frac{M_Z^2\, c_w^2}{s - M_Z^2}\,
\Big[2\, M_Z^2\, \left(6\, s_w^4 - 15\, s_w^2 + 8 \right) + s\, \left(6\, s_w^2 - 5\right)\Big]\,
\mathcal C_0(s,0,M_Z^2,M_W^2,M_W^2,M_W^2) \Bigg]
\nonumber \\
&& \hspace{-20mm}
+ \frac{3 \chi \,(c_w^2 - s_w^2)\,}{s - M_Z^2}\,
\bigg[M_Z^2 +  s\, \bigg( 2\, M_W^2 \, \mathcal C_0(s,0,M_Z^2,M_W^2,M_W^2,M_W^2) 
+ \frac{M_Z^2}{s-M_Z^2}\, \mathcal D_0(s,M_Z^2,M_W^2,M_W^2) \bigg) \bigg]
\Bigg\} \, ,
\nonumber
\eeqa
\beqa
\Xi^{\Sigma}_{\gamma Z}(p,q)
&=&
\frac{\alpha}{\pi}
\Bigg\{ - \frac{c_w\, M_Z^2}{ s_w} \mathcal B_0(0,M_W^2,M_W^2) + 3\, s\, \chi \, s_w^2\, M_Z^2 \, 
\mathcal C_0(s,0,M_Z^2,M_W^2,M_W^2,M_W^2) \Bigg\} \, ,  \nn \\
\Phi^{\Delta}_{\gamma Z}(p,q)
&=& 
\frac{3\,\alpha\,s\,\chi}{\pi s_w c_w (s-M_H^2)(s-M_Z^2)} 
\bigg\{ 2 \sum_f m_f^2 C_v^f Q_f \bigg[ 2 + 2 \frac{M_Z^2}{s-M_Z^2} \mathcal D_0(s,M_Z^2,m_f^2,m_f^2)  \nn \\
&& \hspace{-20mm}
+ (4 m_f^2 + M_Z^2 - s) \mathcal C_0(s,0,M_Z^2,m_f^2,m_f^2,m_f^2) \bigg]  + M_H^2(1-2 s_w^2) + 2 M_Z^2 (6 s_w^4 - 11 s_w^2 +5) \nn \\
&& \hspace{-20mm}
+ \frac{M_Z^2}{s-M_Z^2} ( M_H^2 (1-2 s_w^2) + 2 M_Z^2 (6 s_w^4 -11 s_w^2 + 5) ) \mathcal D_0(s,M_Z^2,M_W^2,M_W^2) \nn \\
&& \hspace{-20mm}
+ 2 M_W^2 \mathcal C_0(s,0,M_Z^2,M_W^2,M_W^2,M_W^2) ( M_H^2 (1-2 s_w^2) + 2 M_Z^2 (6 s_w^4 - 15 s_w^2 + 8) + 2 s (4 s_w^2-3)) \bigg\} 
\nn \\
\Xi^{\Delta}_{\gamma Z}(p,q)
&=&
\frac{3\,\alpha\,s\, \chi\, c_w}{\pi\, s_w}\, M_Z^2 \bigg\{ \frac{2}{s-M_H^2} \mathcal B_0(0, M_W^2,M_W^2) 
- s_w^2 \mathcal C_0(s,0,M_Z^2,M_W^2,M_W^2,M_W^2) \bigg\}\, .
\eeqa

As for the previous case, we give the decay rate in the simplified limit $\chi \to 0$ which is easily found to be
\beqa
\Gamma(\rho\rightarrow \gamma Z) 
&=&
\frac{9\,m_{\rho}^3}{1024\,\Lambda^2\,\pi} \, \sqrt{1-x_Z} \, \bigg( |\Phi^{\Sigma}_{\gamma Z}|^2(p,q)\,m_{\rho}^4\,(x_Z-4)^2 + 
48 \, Re\, \left\{\Phi^{\Sigma}_{\gamma Z}(p,q)\,\Xi^{\Sigma\,*}_{\gamma Z}(p,q) \, m_{\rho}^2\,(x_Z-4)\right\}
\nonumber \\
&& \hspace{35mm}
- \, 192 \, |\Xi^{\Sigma}_{\gamma Z}|^2(p,q)  \bigg) \, ,
\label{RateRhoGammaZ}
\eeqa
where $Re$ is the symbol for the real part.

\subsection{The $\rho Z Z$ vertex}

\begin{figure}[t]
\centering
\subfigure[]{\includegraphics[scale=.7]{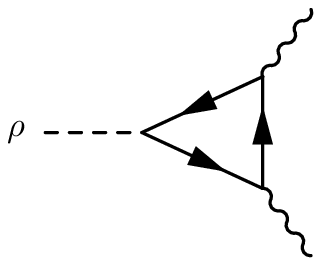}}
\hspace{.2cm}
\subfigure[]{\includegraphics[scale=.7]{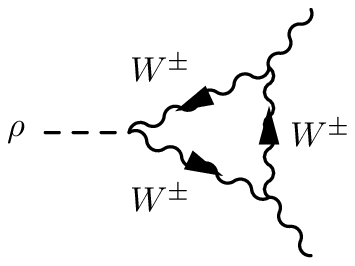}}
\hspace{.2cm}
\subfigure[]{\includegraphics[scale=.7]{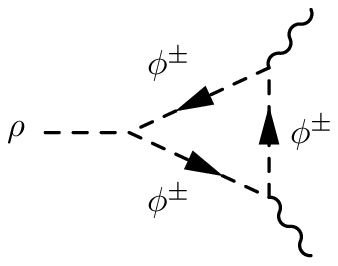}}
\hspace{.2cm}
\subfigure[]{\includegraphics[scale=.7]{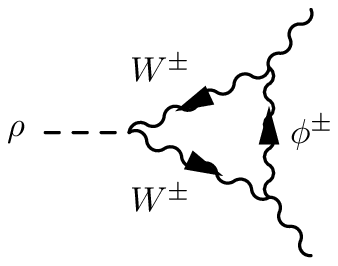}}
\hspace{.2cm}
\subfigure[]{\includegraphics[scale=.7]{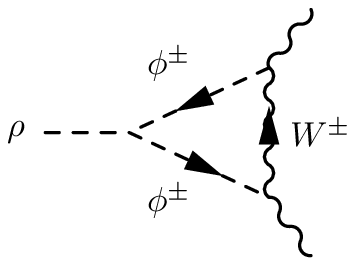}} \\
\hspace{.2cm}
\subfigure[]{\includegraphics[scale=.7]{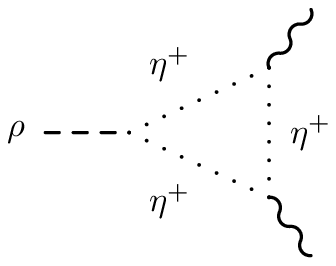}}
\hspace{.2cm}
\subfigure[]{\includegraphics[scale=.7]{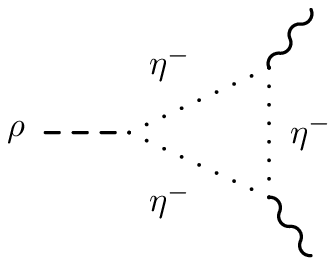}}
\hspace{.2cm}
\subfigure[]{\includegraphics[scale=.7]{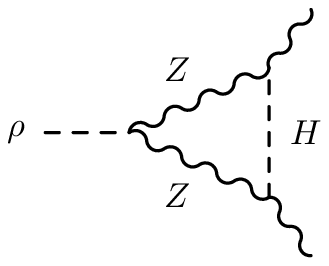}}
\hspace{.2cm}
\subfigure[]{\includegraphics[scale=.7]{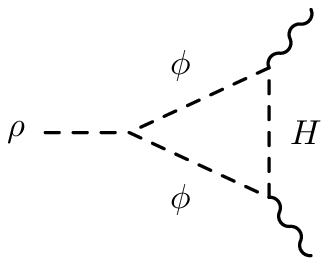}}
\hspace{.2cm}
\subfigure[]{\includegraphics[scale=.7]{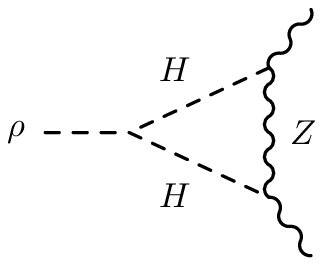}}
\hspace{.2cm}
\subfigure[]{\includegraphics[scale=.7]{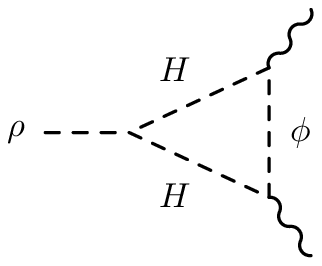}}  
\caption{Amplitudes of triangle topology contributing to the $\rho \gamma\gamma$, $\rho \gamma Z$ and $\rho ZZ$ interactions. They 
include fermion $(F)$, gauge bosons $(B)$ and contributions from the term of improvement (I). Diagrams (a)-(g) contribute to all the 
three channels while (h)-(k) only in the $\rho ZZ$ case.}
\label{figuretriangle}
\end{figure}

\begin{figure}[t]
\centering
\subfigure[]{\includegraphics[scale=.7]{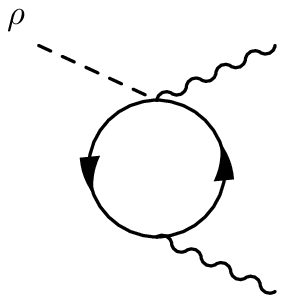}}
\hspace{.2cm}
\subfigure[]{\includegraphics[scale=.7]{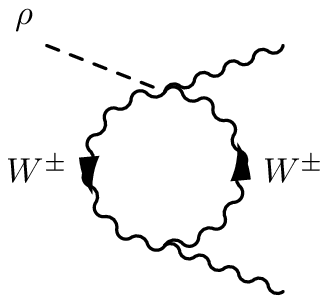}}
\hspace{.2cm}
\subfigure[]{\includegraphics[scale=.7]{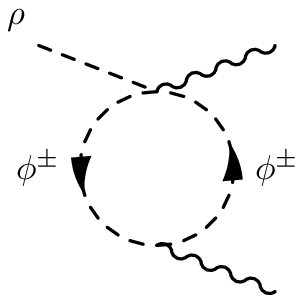}}
\hspace{.2cm}
\subfigure[]{\includegraphics[scale=.7]{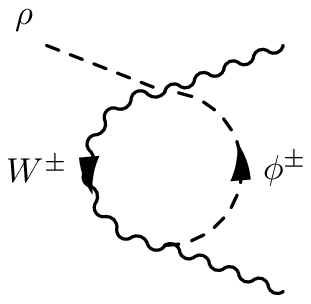}}
\hspace{.2cm}
\subfigure[]{\includegraphics[scale=.7]{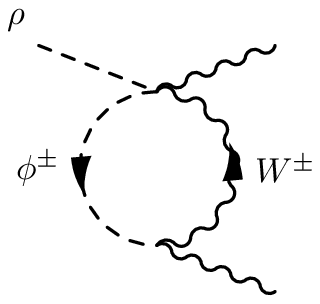}}
\hspace{.2cm}
\subfigure[]{\includegraphics[scale=.7]{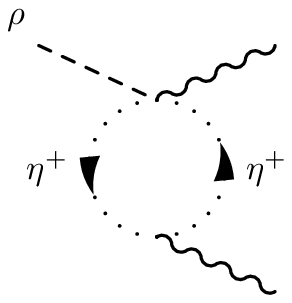}} \\
\subfigure[]{\includegraphics[scale=.7]{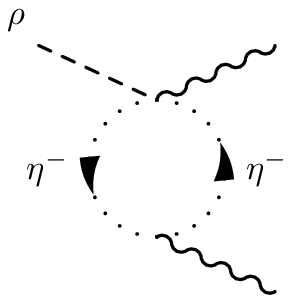}}
\hspace{.2cm}
\subfigure[]{\includegraphics[scale=.7]{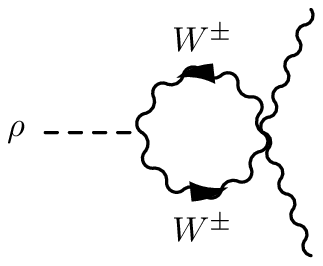}}
\hspace{.2cm}
\subfigure[]{\includegraphics[scale=.7]{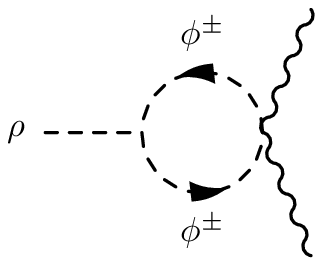}}
\hspace{.2cm}
\subfigure[]{\includegraphics[scale=.7]{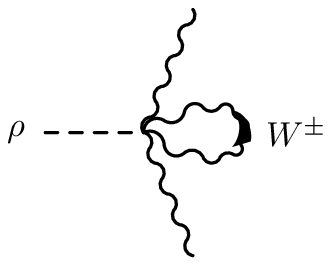}}
\hspace{.2cm}
\subfigure[]{\includegraphics[scale=.7]{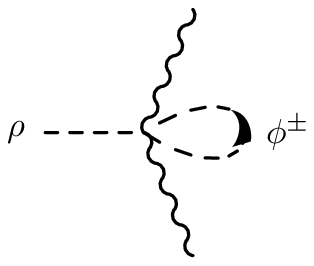}}
\hspace{.2cm}
\subfigure[]{\includegraphics[scale=.7]{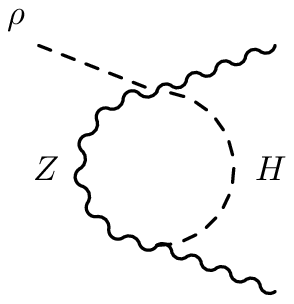}} \\
\hspace{.2cm}
\subfigure[]{\includegraphics[scale=.7]{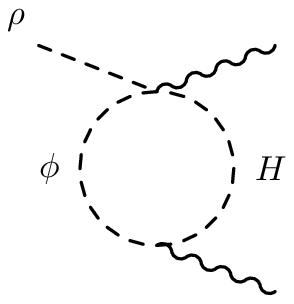}}  
\hspace{.2cm}
\subfigure[]{\includegraphics[scale=.7]{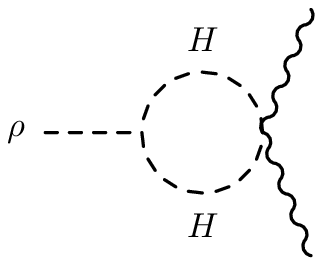}}
\hspace{.2cm}
\subfigure[]{\includegraphics[scale=.7]{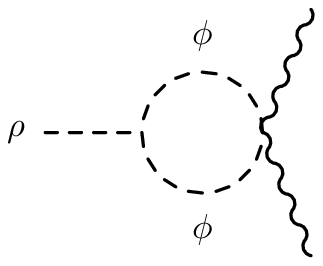}}  
\hspace{.2cm}
\subfigure[]{\includegraphics[scale=.7]{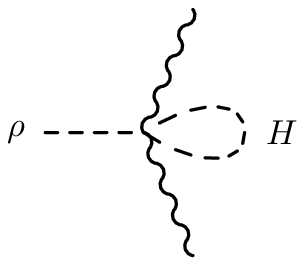}}
\hspace{.2cm}
\subfigure[]{\includegraphics[scale=.7]{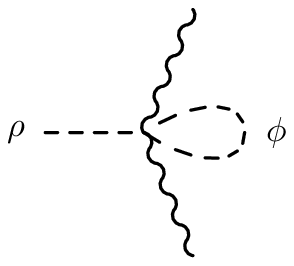}}  
\caption{Bubble and tadpole-like diagrams for $\rho \gamma\gamma$ $\rho \gamma Z $ and $\rho Z Z$. 
Amplitudes (l)-(q) contribute only in the $\rho ZZ$ channel.}
\label{figuretadpole}
\end{figure}

\begin{figure}[t]
\centering
\subfigure[]{\includegraphics[scale=.7]{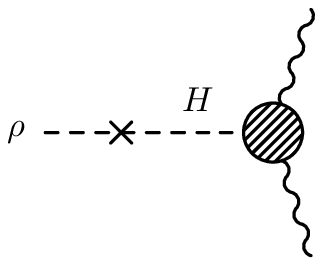}}
\hspace{.2cm}
\subfigure[]{\includegraphics[scale=.7]{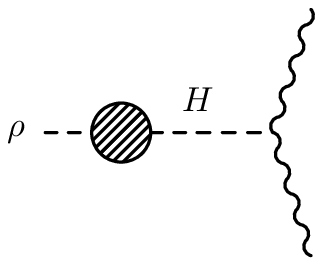}}
\hspace{.2cm}
\subfigure[]{\includegraphics[scale=.7]{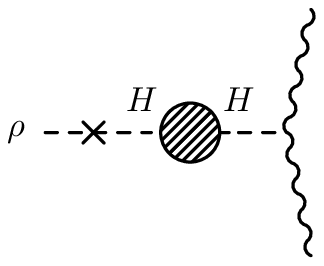}}
\caption{External leg corrections. Diagrams (b) and (c) appear only in the $\rho Z Z$ sector.}
\label{figuremix}
\end{figure}

The expression for the $\Gamma^{\alpha\beta}_{Z Z}$ vertex (Figs. \ref{figuretriangle},\ref{figuretadpole},\ref {figuremix}) defining 
the $\rho ZZ$ interaction is presented here in the kinematical limit given by $k^2 = (p+q)^2 = s$, $p^2 = q^2 = M_Z^2$ with two 
on-shell $Z$ bosons. The completely cut correlator takes contributions from a fermion sector, a $W$ gauge boson sector, a $Z-H$ 
sector together with a term of improvement.
There is also an external leg correction $\Delta^{\alpha \beta}$ on the dilaton line which is much more involved than in the previous 
cases because there are contributions coming from the minimal EMT and from the improved EMT . \\
At one loop order we have 
\bea \label{1loopZZ}
&&
\Gamma_{Z Z}^{\alpha\beta}(p,q)
\equiv 
\frac{i}{\Lambda}\, \left[ \mathcal A^{\alpha\beta}(p,q) + \Sigma^{\alpha\beta}(p,q)+ \Delta^{\alpha\beta}(p,q) \right] \nn \\
&& = 
\frac{i}{\Lambda}\,
\Bigg\{
\left[\left(\frac{s}{2} - M^2_Z \right)\, \eta^{\alpha\beta} - q^\alpha\,p^\beta \right]\,
\left( \mathcal A_{ZZ} + \Phi^{\Sigma}_{ZZ}(p,q) + \Phi^{\Delta}_{ZZ}(p,q) \right)
+ \eta^{\alpha\beta}\, \left(\Xi^{\Sigma}_{ZZ}(p,q) + \Xi^{\Delta}_{ZZ}(p,q) \right) \Bigg\} ,
\eea
where again $\Sigma$ stands for the completely cut vertex and $\Delta$ for the external leg corrections and
we have introduced for convenience the separation
\bea
\Phi^{\Sigma}_{ZZ}(p,q) &=& \Phi^{F}_{ZZ}(p,q) + \Phi^{W}_{ZZ}(p,q) + \Phi^{ZH}_{ZZ}(p,q) + \Phi^{I}_{ZZ}(p,q) \,, \nn \\
\Xi^{\Sigma}_{ZZ}(p,q) &=& \Xi^{F}_{ZZ}(p,q) + \Xi^{W}_{ZZ}(p,q) + \Xi^{ZH}_{ZZ}(p,q) + \Xi^{I}_{ZZ}(p,q)\, .
\eea
The form factors are given in Appendix \ref{VZZ}, while here we report only the purely anomalous contribution
\beq
\mathcal A_{ZZ} = \frac{\alpha}{6 \pi c_w^2 s_w^2}\,
\left\{ -\sum_{f} \left({C_a^f}^2+{C_v^f}^2\right) + \frac{60\,s_w^6 - 148\,s_w^2 + 81}{4} - \frac{7}{4} 
+ 18\,\chi\,\left[ 1 - 2\,s_w^2\,c_w^2 \right]\right\}\, .
\eeq
Finally, we give the decay rate expression for the $\rho \to  Z Z$ process. 
At leading order it can be computed from the tree level amplitude
\beqa
\mathcal M^{\alpha\beta}(\rho \rightarrow ZZ) 
&=& 
\frac{2}{\Lambda}\,M_Z^2\, \eta^{\alpha\beta} \, , 
\eeqa
and it is given by  
\beqa
\Gamma(\rho \rightarrow ZZ) 
&=&
\frac{ m_{\rho}^3}{32\,\pi \Lambda^2} \, (1-x_Z)^{1/2}\, \left[ 1 - x_Z + \frac{3}{4}\,x_Z^2 \right].
\label{PhiZZ}
\eeqa
Including the one-loop corrections defined in Eq.(\ref{1loopZZ}), one gets the decay rate at next-to-leading order
\beqa
 \Gamma(\rho\rightarrow ZZ) 
&=&
\frac{m_{\rho}^3}{32\,\pi\,\Lambda^2} \,\sqrt{1-x_Z}\, \bigg\{ 1 - x_Z + \frac{3}{4}\,x_Z^2 
+ \frac{3}{x_Z} \, \bigg[4\, Re\, \{\Phi^{\Sigma}_{ZZ}(p,q)\}(1-x_Z + \frac{3}{4}\,x_Z^2) 
\nn \\
&-& 
Re\, \{\Xi^{\Sigma}_{ZZ}(p,q)\}\,m_{\rho}^2\, \left( \frac{3}{4}\,x_Z^3 - \frac{3}{2}\,x_Z^2 \right) \bigg] \bigg\}\, .
\eeqa
%
%
%
\subsection{Renormalization of dilaton interactions in the broken electroweak phase}
\label{renorm}

In this section we address the renormalization properties of the correlation functions given above. Although the 
proof is quite cumbersome, one can check, from our previous results, that the 1-loop renormalization of the Standard Model 
Lagrangian is sufficient to cancel all the singularities in the cut 
vertices independently of whether the Higgs is conformally coupled or not. Concerning the uncut vertices, instead, the term of 
improvement plays a significant role in the determination of Green functions which are ultraviolet finite. In particular such a term 
has to appear with $\chi=1/6$ in order to guarantee the cancellation of a singularity present in the 1-loop 2-point function 
describing the Higgs dilaton mixing ($\Sigma_{\rho H}$).  
The problem arises only in the $\Gamma^{\alpha\beta}_{ZZ}$ correlator, where the $\Sigma_{\rho H}$ two-point function is present as 
an external leg correction on the dilaton line. 

The finite parts of the counterterms are determined in the on-shell renormalization scheme, which is widely used in the electroweak 
theory. In this scheme the renormalization conditions are fixed in terms of the physical parameters to all orders in perturbation 
theory and the wave-function normalizations of the fields are obtained by requiring a unit residue of the full 2-point functions on 
the physical particle poles.

From the counterterm Lagrangian we compute the corresponding counterterm to the trace of the EMT. As we have already mentioned, one can also verify from the 
explicit computation that the terms of improvement, in the conformally coupled case, are necessary to renormalize the vertices 
containing an intermediate scalar with an external bilinear mixing (dilaton/Higgs).
The counterterm vertices for the correlators with a dilaton insertion are
\beqa
\delta [\rho \gamma \gamma]^{\alpha\beta}  &=& 0
\\
\delta [\rho \gamma Z]^{\alpha\beta}  &=&  - \frac{i}{\Lambda}  \delta Z_{Z\gamma} \, M_Z^2  \, \eta^{\alpha\beta}  \, , 
\\
\delta [\rho Z Z]^{\alpha\beta}  &=&  - 2 \frac{i}{\Lambda}  (M_Z^2 \, \delta Z_{ZZ} + \delta M^2_Z)  \, \eta^{\alpha\beta} \, , 
\eeqa
where the counterterm coefficients are defined in terms of the 2-point functions of the fundamental fields as
\beqa
\delta Z_{Z \gamma} = 2 \frac{\Sigma_T^{\gamma Z}(0)}{M_Z^2} \,, \quad \delta Z_{ZZ} = - Re \frac{\partial 
\Sigma_T^{ZZ}(k^2)}{\partial k^2} \bigg |_{k^2 = M_Z^2} \,, \quad
\delta M_Z^2 = Re \, \Sigma_T^{ZZ}(M_Z^2) \,,
\eeqa
and are defined in Appendix \ref{SigmaSM}.
It follows then that the $\rho \gamma \gamma$ interaction must be finite, as one can find by a direct inspection of the $\Gamma^{\alpha\beta}_{\gamma \gamma}$ vertex, while the others require the subtraction of their divergences.

These counterterms are sufficient to remove the divergences of the completely cut graphs which do 
not contain a bilinear mixing, once we set on-shell  the external gauge lines. This occurs both for those diagrams which do not 
involve the terms of improvement and for those involving $T_I$. 
Regarding those contributions which involve the bilinear mixing on the external dilaton line, we encounter two different situations. \\
In the $\rho \gamma Z$ vertex the insertion of the bilinear mixing $\rho H$ generates a reducible diagram of the form Higgs/photon/Z 
whose renormalization is guaranteed, within the Standard Model, by the use of the Higgs/photon/Z counterterm
\beqa
\delta [H\gamma Z]^{\alpha\beta} = \frac{e \, M_Z}{2 s_w c_w} \delta Z_{Z\gamma}\, \eta^{\alpha\beta} \,.
\eeqa
As a last case, we discuss the contribution to $\rho ZZ$ coming from the bilinear mixing, already mentioned above. The corrections on 
the dilaton line involve the dilaton/Higgs mixing $\Sigma_{\rho H}$, the Higgs self-energy $\Sigma_{HH}$ and the term of improvement 
$\Delta^{\alpha\beta}_{I\,,HZZ}$, which introduces the Higgs/Z/Z vertex (or $HZZ$) of the Standard Model. 
The Higgs self-energy and the $HZZ$ vertex, in the Standard Model, are renormalized with the usual counterterms
\beqa
\delta [HH](k^2) 
&=& 
(\delta Z_H \, k^2 - M_H^2 \delta Z_H - \delta M_H^2) \, , 
\\
\delta [HZZ]^{\alpha\beta} 
&=& 
\frac{e \, M_Z}{s_w \, c_w} \bigg[ 1 + \delta Z_e + \frac{2 s_w^2 
- c_w^2}{c_w^2} \frac{\delta s_w}{s_w} + \frac{1}{2} \frac{\delta M_W^2}{M_W^2} + \frac{1}{2} \delta Z_H + \delta Z_{ZZ}  \bigg] 
\, \eta^{\alpha\beta} \,,
\eeqa
where
\bea
&& 
\delta Z_H = 
- Re \frac{\partial \Sigma_{HH}(k^2)}{\partial k^2} \bigg|_{k^2=M_H^2}\, ,\quad \delta M_H^2 = Re \Sigma_{HH}(M_H^2) \, , \quad
\delta Z_e = - \frac{1}{2} \delta Z_{\gamma \gamma} + \frac{s_w}{2 c_w} \delta Z_{Z \gamma} \,, \nn \\
&& \delta s_w = - \frac{c_w^2}{2 s_w} \left( \frac{\delta M_W^2}{M_W^2} - \frac{\delta M_Z^2}{M_Z^2} \right) \,, \quad \delta M_W^2 = 
Re \Sigma^{WW}_T(M_W^2) \,, \quad 
\delta Z_{\gamma \gamma} = - \frac{\partial \Sigma_T^{\gamma \gamma}(k^2)}{\partial k^2} \bigg|_{k^2 = 0} \,.
\eea
The self-energy $\Sigma_{\rho H}$ is defined by the minimal contribution generated by ${{T_{Min}}^\mu}_\mu$ 
and by a second term derived from ${{T_I}^\mu}_\mu$.
This second term, with the conformal coupling $\chi = \frac{1}{6}$, is necessary in order to ensure the renormalizability 
of the dilaton/Higgs mixing.
In fact, the use of the minimal EMT in the computation of this self-energy involves a divergence of the form
\beqa
\delta [\rho H]_{Min} = - 4 \frac{i}{\Lambda}\, \delta t \, , \label{CThH}
\eeqa
with $\delta t$ fixed by the condition of cancellation of the Higgs tadpole $T_{ad}$ ($\delta t + T_{ad} = 0$).
A simple analysis of the divergences in $\Sigma_{Min, \, \rho H}$ shows that the counterterm given in Eq. (\ref{CThH}) is not 
sufficient to remove all the singularities of this correlator unless we also include the renormalization constant provided by the term of improvement 
which is given by
\beqa
\delta [\rho H]_{I}(k) = - \frac{i}{\Lambda} 6 \, \chi \, v \bigg[ \delta v + \frac{1}{2} \delta Z_H \bigg]  k^2\,, \qquad 
\qquad \text{with} \quad \chi = \frac{1}{6} \,,
\eeqa
and
\bea
\delta v = v \bigg( \frac{1}{2} \frac{\delta M_W^2}{M_W^2} + \frac{\delta s_w}{s_w} - \delta Z_e \bigg) \,.
\eea
One can show explicitly that this counterterm indeed ensures the finiteness of $\Sigma_{\rho H}$.



\begin{figure}[t]
\centering
\subfigure[]{\includegraphics[scale=.7]{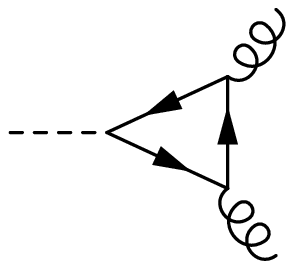}\label{QCD_NLOa}}
\hspace{.5cm}
\subfigure[]{\includegraphics[scale=.7]{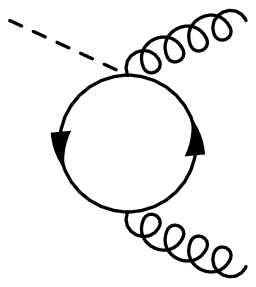}\label{QCD_NLOb}}
\hspace{.5cm}
\subfigure[]{\includegraphics[scale=.7]{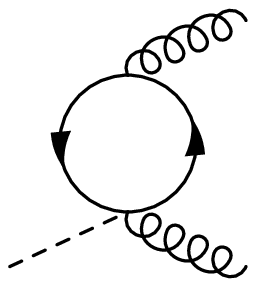}\label{QCD_NLOc}}
\hspace{.5cm}
\subfigure[]{\includegraphics[scale=.7]{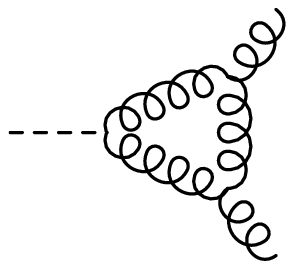}\label{QCD_NLOd}}
\hspace{.5cm}
\subfigure[]{\includegraphics[scale=.7]{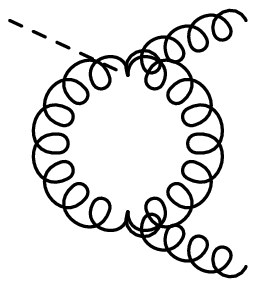}\label{QCD_NLOe}}
\hspace{.5cm}
\subfigure[]{\includegraphics[scale=.7]{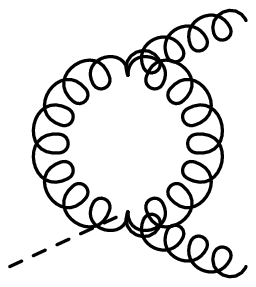}\label{QCD_NLOf}}\\
\hspace{.5cm}
\subfigure[]{\includegraphics[scale=.7]{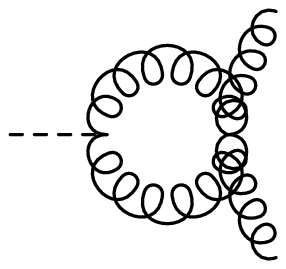}\label{QCD_NLOg}}
\hspace{.5cm}
\subfigure[]{\includegraphics[scale=.7]{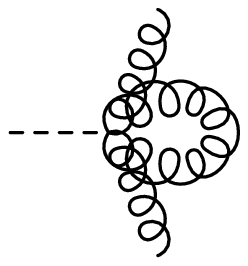}\label{QCD_NLOh}}
\hspace{.5cm}
\subfigure[]{\includegraphics[scale=.7]{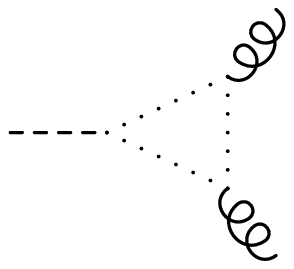}\label{QCD_NLOi}}
\hspace{.5cm}
\subfigure[]{\includegraphics[scale=.7]{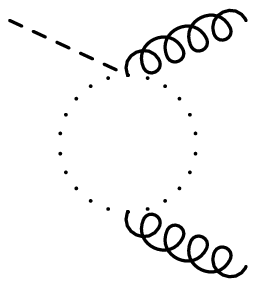}\label{QCD_NLOl}}
\hspace{.5cm}
\subfigure[]{\includegraphics[scale=.7]{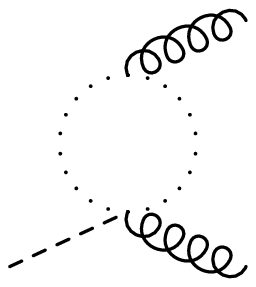}\label{QCD_NLOm}}
\caption{QCD vertices at next-to-leading order. In the on-shell gluon case only diagram (a)contributes.}
\label{QCD_NLO}
\end{figure}
%

\section{The off-shell dilaton-gluon-gluon vertex in QCD}

After a discussion of the leading corrections to the vertices involving one dilaton and two electroweak currents we investigate the 
interaction of a dilaton and two gluons beyond leading order, giving the expression of the full off-shell vertex. The corresponding 
interaction with two on-shell gluons has been computed in \cite{Giudice:2000av} and is simply given by the contributions of the 
anomaly and of the quark loop. We will come back to rediscuss the on-shell case in the second part of this work, where we will 
stress some specific perturbative features of this interaction.  

We show in Fig. \ref{QCD_NLO} a list of the NLO QCD contributions to dilaton interactions. As we have just mentioned, in the 
two-gluon on-shell case one can show by an explicit computation that each of these contributions vanishes, except for diagram $(a)$, 
which is nonzero when a massive fermion runs in the loop. For this specific reason, in the parton model, the production of the 
dilaton in $pp$ collisions at the LHC is mediated by the diagram of gluon fusion, which involves a top quark in a loop. 

We find convenient to express the result of the off-shell $\Gamma^{\alpha\beta}_{g g}$ vertex in the form 
\beqa \label{OffShellQCD}
\Gamma_{g g}^{\alpha \beta}(p,q) = 
\frac{i}{\Lambda}\, \bigg\{ A^{00}(p,q) \eta^{\alpha \beta} + A^{11}(p,q) p^{\alpha} p^{\beta} + A^{22}(p,q) q^{\alpha} q^{\beta} 
+ A^{12}(p,q) p^{\alpha}q^{\beta} + A^{21}(p,q) q^{\alpha}p^{\beta}\bigg\} \, ,
\eeqa
where $A^{ij}(p,q) = A^{ij}_g(p,q) + A^{ij}_q(p,q)$ which are diagonal ($\propto \delta^{a b}$) in color space.\\
After an explicit computation, we find
\beqa
A^{00}_g(p,q) 
&=& 
-\delta_{ab}\,\frac{g^2 \, N_C}{16 \pi^2} \bigg\{   
2 \left( p^2 +  q^2 + \frac{11}{3} p \cdot q \right) 
+ (p^2-q^2) \bigg[ \mathcal B_0(p^2,0,0) - \mathcal B_0(q^2,0,0)\bigg] 
\nonumber \\
&& 
+ \left(p^4 + q^4 - 2(p^2 + q^2) p \cdot q -6 p^2 q^2 \right) \mathcal C_0((p+q)^2, p^2, q^2,0,0,0)
\bigg\} \, , 
\nonumber 
\eeqa
\beqa
A^{11}_g(p,q) = A^{22}_g(q,p)
&=&   
\delta_{ab} \, \frac{g^2 \, N_C \,}{16\,\pi^2 }  \, 
\bigg\{ 2 + \frac{1}{p \cdot q^2 - p^2 \, q^2} \, 
\bigg[ (p+q)^2 \, p \cdot q \, \mathcal B_0((p+q)^2,0,0)
\nonumber \\
&& 
- p^2 \, (q^2 + p \cdot q) \, \mathcal B_0(p^2,0,0)  - (2 p \cdot q^2 - p^2 \, q^2 + p \cdot q \, q^2) \mathcal B_0(q^2,0,0) 
\nonumber \\
&& 
+ \left( p^2 \, q^2 ( 5 q^2 - p^2) + 2 p \cdot q^2 (p^2 + p \cdot q - 2 \, q^2 ) \right) \, 
\mathcal C_0((p+q)^2, p^2, q^2,0,0,0) \bigg] \bigg\}  \, , 
\nonumber 
\eeqa
\beqa
A^{12}_g(p,q) 
&=& 
\delta_{ab}\, \frac{g^2 \, N_C}{4 \pi^2} \, p \cdot q \, \mathcal C_0((p+q)^2, p^2, q^2,0,0,0)\, ,
\nonumber 
\eeqa
\beqa
A^{21}_g(p,q) 
&=& 
\delta_{ab}\, \frac{g^2 \, N_C}{24\,\pi^2 }  \, 
\bigg\{11 + \frac{3}{2}\,\frac{1}{p \cdot q^2 - p^2 \, q^2} (p^2 + q^2) \, \bigg[(p^2 + p \cdot q)\mathcal B_0(p^2,0,0) 
\nonumber \\
&& 
+ (q^2 + p \cdot q) \mathcal B_0(q^2,0,0) - (p+q)^2  \mathcal B_0((p+q)^2,0,0)  
\nonumber \\
&& 
- \left( p \cdot q (p^2 + 4 p \cdot q + q^2) - 2 \, p^2 \, q^2 \right) \mathcal C_0((p+q)^2, p^2, q^2,0,0,0) \bigg]\bigg\} \, ,
\nonumber 
\eeqa
\bea
A^{00}_q(p,q) &=&
\delta_{ab} \frac{g^2}{8 \pi^2} \sum_{i=1}^{n_f} \bigg\{ \frac{2}{3} p\cdot q - 2\, m_i^2 
+ \frac{m_i^2}{p \cdot q^2 - p^2 q^2} \bigg[
p^2 \left(p \cdot q +q^2\right)  \mathcal B_0(p^2,m_i^2,m_i^2 ) \nn \\
&+&
q^2 \left(p^2+p \cdot q \right) \mathcal B_0(q^2,m_i^2,m_i^2)  -
\left(p^2 \left(p \cdot q+2 q^2\right) + p \cdot q \, q^2\right) \mathcal B_0( (p+q)^2,m_i^2,m_i^2) \nn \\
&-&
\left(p^2 q^2 \left(p^2+q^2 -4 m_i^2 \right)+4 m_i^2 \, p \cdot q^2 +4 p^2 \, q^2 \, p \cdot q -2 \, p \cdot q^3\right)
  \mathcal C_0 ((p+q)^2,p^2,q^2,m_i^2,m_i^2,m_i^2) 
\bigg] \bigg\} , \nn
\eea
%
\bea
A^{11}_q(p,q) &=& A^{22}_q(q,p) =
\delta_{ab} \frac{g^2}{ 8\,\pi^2} \sum_{i=1}^{n_f} \frac{2 \, m_i^2 \, q^2}{ p \cdot q^2 - p^2 q^2} \bigg\{ -2 + \frac{1}{p \cdot q^2 - p^2 q^2} \bigg[
\left(q^2 \left(p^2+3 \, p \cdot q \right) \right. \nn \\
&+& \left. 2 \, p \cdot q^2\right) \mathcal B_0(q^2,m_i^2,m_i^2)
+ \left(p^2 \left(3 \, p \cdot q+q^2\right)+2 p \cdot q^2\right)  \mathcal B_0(p^2,m_i^2,m_i^2) \nn \\
&-& \left(p^2 \left(3 \, p \cdot q+2 q^2\right) + p \cdot q \left(4 \, p \cdot q+3 q^2\right)\right) \mathcal B_0((p+q)^2,m_i^2,m_i^2) \nn \\
&-& \left(2 \, p \cdot q^2 \left(2 m_i^2+p^2+q^2\right) + p^2 q^2 \left(p^2+q^2-4 m_i^2 \right)  \right.) \nn \\
&+& \left. 4 p^2 q^2 \, p \cdot q +2 \, p \cdot q ^3\right)
   \mathcal C_0((p+q)^2,p^2,q^2,m_i^2,m_i^2,m_i^2)
\bigg] \bigg\} , \nn
\eea
\bea
A^{12}_q(p,q) &=&
\delta_{ab} \frac{g^2}{ 8\,\pi^2} \sum_{i=1}^{n_f} \frac{2 \, m_i^2 \, p \cdot q^2}{ p \cdot q^2 - p^2 q^2} \bigg\{ 2 + \frac{1}{p \cdot q^2 - p^2 q^2} \bigg[
-\left(q^2 \left(p^2+3 \, p \cdot q \right)+ 2 \, p \cdot q^2\right) \mathcal B_0(q^2,m_i^2,m_i^2) \nn \\
&-&
\left(p^2 \left(3 \, p \cdot q +q^2\right)+ 2 \, p \cdot q^2\right) \mathcal B_0(p^2,m_i^2,m_i^2) +
\left(p^2 \left(3 \, p \cdot q +2 q^2\right)+p \cdot q \left(4 \, p \cdot q +3 q^2 \right)\right) \nn \\
&\times& \mathcal B_0( (p+q)^2,m_i^2,m_i^2)+
\left(2 \, p \cdot q^2 \left(2 m_i^2+p^2+q^2\right)+p^2 q^2 \left(p^2+q^2 -4 m_i^2 \right) \right. \nn \\
&+& \left.  4 \, p^2 \, q^2 \, p \cdot q +2\, p \cdot q^3\right)
 \mathcal C_0((p+q)^2,p^2,q^2,m_i^2,m_i^2,m_i^2)
\bigg] \bigg\} , \nn 
\eea
\bea
A^{21}_q(p,q)  &=& \delta_{ab} \frac{g^2}{ 8\,\pi^2} \sum_{i=1}^{n_f} \bigg\{ - \frac{2}{3} + \frac{2 \, m_i^2 \, p \cdot q}{p \cdot q - p^2 q^2} + \frac{m_i^2}{(p \cdot q -p^2 q^2)^2} \bigg[
- p^2 \left(q^2 \left(2 p^2+3 \, p \cdot q \right)+p \cdot q ^2\right)  \nn \\ 
&\times& \mathcal B_0(p^2, m_i^2,m_i^2) 
- q^2 \left(p^2 \left(3 \, p \cdot q+2 q^2\right)+p \cdot q^2\right) \mathcal B_0( q^2, m_i^2,m_i^2) 
+ \left(2 p^4 q^2+p^2 \left(6 \, p \cdot q \, q^2 \right. \right. \nn \\ 
&+& \left. \left. p \cdot q^2+2 q^4\right)+p \cdot q^2 q^2\right) \mathcal B_0( (p+q)^2,m_i^2,m_i^2)
+ p \cdot q \left(p^2 q^2 \left(3 p^2+3 q^2 - 4 m_i^2\right) \right. \nn \\
&+& \left. 4 \,   m_i^2 \, p \cdot q^2
 +8 p^2 \, q^2 \, p \cdot q -2 \, p \cdot q ^3\right)
   \mathcal C_0( (p+q)^2,p^2,q^2,m_i^2,m_i^2,m_i^2)
\bigg] \bigg\}, 
\eea
where $N_C$ is the number of colors, $n_f$ is the number of flavor and $m_i$ the mass of the quark.
In the on-shell gluon case, Eq.(\ref{OffShellQCD}) reproduces the same interaction responsible for Higgs production at LHC augmented by an anomaly term.
This is given by
\beqa \label{OnShellQCD}
\Gamma^{\alpha\beta}_{gg}(p,q) =  
\frac{i}{\Lambda} \Phi(s) \, u^{\alpha\beta}(p,q) \, ,
\eeqa
with $u^{\alpha\beta}(p,q)$ defined in Eq.(\ref{utensor}), and with the gluon/quark contributions included in the $\Phi(s)$ form factor ($s = k^2 = (p+q)^2$) 
\beqa \label{OnShellPhi}
\Phi(s) 
= 
- \delta^{ab} \frac{g^2}{24\,\pi^2} \, \bigg\{
 \, \left(11\, N_C - 2\, n_f \right) + 12 \, \sum_{i=1}^{n_f} m_i^2 \, 
\bigg[ \frac{1}{s} \, - \, \frac{1} {2 }\mathcal C_0 (s, 0, 0, m_i^2, m_i^2, m_i^2) \bigg(1-\frac{4 m_i^2}{ s}\bigg) \bigg]
\bigg\} \, ,
\eeqa
where the first mass independent terms represent the contribution of the anomaly, while the others are the explicit mass corrections. \\ 
The decay rate of a dilaton in two gluons can be evaluated from the on-shell limit in Eq.(\ref{OnShellQCD}) and it is given by
\beqa
\Gamma(\rho \rightarrow gg) 
&=&
\frac{\alpha_s^2\,m_\rho^3}{32\,\pi^3 \Lambda^2} \, \bigg| \beta_{QCD} + x_t\left[1 + (1-x_t)\,f(x_t) \right] \bigg|^2 \,,
\label{Phigg}
\eeqa
where we have taken the top quark as the only massive fermion and $x_i$ and $f(x_i)$ are defined in Eq. (\ref{x}) and Eq. (\ref{fx}) respectively. 
Moreover we have set
$\beta_{QCD} = 11 N_C/3 - 2 \, n_f/3$ for the QCD $\beta$ function.

\section{Non-gravitational dilatons from scale-invariant extensions of the Standard Model}
\label{NonGrav} 

As we have pointed out in the introduction, a dilaton may appear in the spectrum of different extensions of the Standard Model not 
only as a result of the compactification of extra spacetime dimensions, but also as an effective state, related to the breaking of a 
dilatation symmetry. In this respect, notice that in its actual formulation 
the Standard Model is not scale-invariant, but can be such, at classical level, if we slightly modify the scalar potential with the 
introduction of a dynamical field $\Sigma$ that 
allows to restore this symmetry and acquires a vacuum expectation value. This task is accomplished by the replacement of every 
dimensionfull parameter $m$ according to $m \rightarrow m \frac{\Sigma}{\Lambda}$, where $\Lambda$ is the classical conformal 
breaking scale. 
In the case of the Standard Model, classical scale invariance can be easily accomodated with a simple change of the scalar potential. 

This is defined, obviously, modulo a constant, therefore we may consider, for instance, two equivalent choices 
\beqa
V_1(H, H^\dagger)&=& - \mu^2 H^\dagger H +\lambda(H^\dagger H)^2 =
\lambda \left( H^\dagger H - \frac{\mu^2}{2\lambda}\right)^2 - \frac{\mu^4}{4 \lambda}\nonumber \\
V_2(H,H^\dagger)&=&\lambda \left( H^\dagger H - \frac{\mu^2}{2\lambda}\right)^2
\eeqa
which gives two {\em different} scale-invariant extensions 
\beqa
V_1(H,H^\dagger, \Sigma)&=&- \frac{\mu^2\Sigma^2}{\Lambda^2} H^\dagger H +\lambda(H^\dagger H)^2 \nonumber \\
V_2(H,H^\dagger, \Sigma)&=& \lambda \left( H^\dagger H - \frac{\mu^2\Sigma^2}{2\lambda \Lambda^2}\right)^2 \,,
\eeqa 
where $H$ is the Higgs doublet, $\lambda$ is its dimensionless coupling constant, while $\mu$ has the dimension of a mass and, 
therefore, is the only term involved in the scale invariant extension. More details of this analysis can be found in appendix \ref{classical}.\\
The invariance of the potential under the addition of constant terms, typical of any Lagrangian, is lifted once we 
require the presence of a dilatation symmetry. Only the second choice $(V_2)$ guarantees the existence of a stable ground state 
characterized by a spontaneously 
broken phase. In $V_2$ we parameterize the Higgs, as usual, around the electroweak vev $v$ as in Eq.(\ref{Higgsparam}), 
and indicate with $\Lambda$ the vev of the dilaton field $\Sigma = \Lambda + \rho$, 
and we have set $\phi^+ = \phi = 0$ in the unitary gauge. \\
The potential $V_2$ has a massless mode due to the existence of a flat direction. 
Performing a diagonalization of the mass matrix we define the two mass eigenstates $\rho_0$ and $h_0$, which are given by 
\beq
 \left( \begin{array}{c}
 {\rho_0}\\
  h_0 \\
  \end{array} \right)
 =\left( \begin{array}{cc}
\cos\alpha & \sin\alpha \\
-\sin\alpha & \cos\alpha  \\
 \end{array} \right)
 \left( \begin{array}{c}
  \rho\\
 {h} \\
  \end{array} \right)
\eeq
with 
\beq
\cos\alpha=\frac{1}{\sqrt{1 + v^2/\Lambda^2}}\qquad \qquad  \sin\alpha=\frac{1}{\sqrt{1 + \Lambda^2/v^2}}.
\eeq
We denote with ${\rho_0}$ the massless dilaton generated by this potential, while 
$h_0$ will describe a massive scalar, interpreted as a new Higgs field, whose mass is given by  
\beq 
m_{h_0}^2= 2\lambda v^2 \left( 1 +\frac{v^2}{\Lambda^2}\right) \qquad \textrm{with} \qquad v^2=\frac{\mu^2}{\lambda},
\eeq
and with $m_h^2=2 \lambda v^2$ being the mass of the Standard Model Higgs.
The Higgs mass, in  this case, is corrected by the new scale of the spontaneous breaking of the dilatation symmetry ($\Lambda$), 
which remains a free parameter. 
 
The vacuum degeneracy of the scale-invariant model  can be lifted by the introduction of 
extra (explicit breaking) terms which give a small mass to the dilaton field.
To remove such degeneracy, one can introduce, for instance, the term
\beq
\mathcal{L}_{break} 
= \frac{1}{2} m_{\rho}^2 {\rho}^2 + \frac{1}{3!}\, {m_{\rho}^2} \frac{{\rho}^3}{\Lambda} + \dots \, ,
\eeq
where $m_{\rho}$ represents the dilaton mass.

It is clear that in this approach the coupling of the dilaton to the anomaly has to be added by hand.
The obvious question to address, at this point, is if one can identify in the effective action of the Standard Model 
an effective state which may interpolate between the dilatation current of the same model and the final state with two
neutral currents, for example with two photons. The role of the following sections will be to show 
rigorously that such a state can be identified in ordinary perturbation theory in the form of an anomaly pole.

We will interpret this scalar exchange as a composite state whose interactions with the rest of the Standard Model are 
defined by the conditions of scale and gauge invariance. In this respect, the Standard Model Lagrangian, enlarged by the introduction 
of a potential of the form $V_2(H,H^\dagger,\Sigma)$, which is expected to capture the dynamics of this pseudo-Goldstone mode, could
take the role of a workable model useful for a phenomenological analysis.  
We will show rigorously that this state couples to the conformal anomaly by a direct analysis of the $J_DVV$ correlator, 
in the form of an anomaly pole, with $J_D$ and $V$ being the dilatation and a vector current respectively.
Usual polology arguments support the fact that a pole  in a correlation function is there to indicate that a specific state can be created by 
a field operator in the Lagrangian of the theory, or, alternatively, as a composite particle of the same elementary fields.  

Obviously, a perturbative hint of the existence of such intermediate state does not correspond to a complete 
description of the state, in the same way as the discovery of an anomaly pole in the $AVV$ correlator of QCD (with $A$ being the 
axial current) is not equivalent to a proof of the existence of the pion. Nevertheless, massless poles extracted from the perturbative effective action do not appear for no reasons, 
and their infrared couplings should trigger further phenomenological interest. 

\subsection{The $J_DVV$ and $TVV$ vertices}

This effective degree of freedom emerges both from the spectral analysis of the $TVV$ \cite{Giannotti:2008cv, Armillis:2009pq} and, 
as we are now going to show, of the  $J_D VV$ 
vertices, being the two vertices closely related. 
We recall that the dilatation current can be defined as 
\beq
J_D^\mu(z)= z_\nu T^{\mu \nu}(z) \qquad \textrm{with}  \qquad \partial\cdot J_D = {T^\mu}_\mu. 
\label{def}
\eeq
The $T^{\mu\nu}$ has to be symmetric and on-shell traceless for a classical scale-invariant theory, and includes, at
quantum level, the contribution from the trace anomaly together with the additional terms describing the explicit breaking of the 
dilatation symmetry. 
The separation between the anomalous and the explicit contributions to the breaking of dilatation symmetry is present 
in all the analysis that we have performed on the $TVV$ vertex in dimensional regularization. 
In this respect, the analogy between these types of correlators and the $AVV$ diagram of the chiral anomaly goes 
quite far, since in the $AVV$ case such a separation has been shown to hold in the Longitudinal/Transverse 
(L/T) solution of the anomalous Ward identities \cite{Knecht:2003xy, Jegerlehner:2005fs, Armillis:2009sm}. 
This has been verified in perturbation theory in the same scheme.

We recall that the $U(1)_A$ current is characterized by an anomaly pole which describes the interaction between the 
Nambu-Goldstone mode, generated by the breaking of the chiral symmetry, and the gauge currents. 
In momentum space this corresponds to the nonlocal vertex 
\beq
\label{AVVpole}
V_{\textrm{anom}}^{\lambda \mu\nu}(k,p,q)=  \frac{k^\lambda}{k^2}\epsilon^{\mu \nu \alpha \beta}p_\alpha q_\beta +...
\eeq
with $k$ being the momentum of the axial-vector current and $p$ and $q$ the momenta of the two photons.
In the equation above, the ellipsis refer to terms which are suppressed at large energy. 
In this regime, this allows to distinguish the operator accounting for the chiral anomaly (i.e. $\square^{-1}$ in coordinate space)
from the contributions due to mass corrections. 
Polology arguments can be used to relate the appearance of such a pole to the pion state 
around the scale of chiral symmetry breaking. 

To identify the corresponding pole in the dilatation current of the $J_D VV$ correlator at zero momentum transfer, one can follow the 
analysis of \cite{Horejsi:1997yn}, where it is shown that the appearance of the trace anomaly is related to the presence of a
superconvergent sum rule in the spectral density of this correlator. At nonzero momentum transfer the derivation of a similar
behaviour can be obtained by an explicit computation of the spectral density of the 
$TVV$ vertex \cite{Giannotti:2008cv} or of the entire correlator, as done for QED and QCD 
\cite{Armillis:2009pq, Armillis:2010qk} and as we will show next.

Using the relation between $J_D^\mu$ and the EMT $T^{\mu\nu}$ we introduce the $J_DVV$ correlator 
\beqa
\Gamma_D^{\mu\alpha\beta}(k,p)
&\equiv& 
\int d^4 z\, d^4 x\, e^{-i k \cdot z + i p \cdot x}\,
\left\langle J^\mu_D(z) V^\alpha(x)V^\beta(0)\right\rangle 
\label{gammagg}
\eeqa
which can be related to the $TVV$ correlator 
\beqa
\Gamma^{\mu\nu\alpha\beta}(k,p)&\equiv& \int d^4 z\, d^4 x\, e^{-i k \cdot z + i p \cdot x}\, 
\left\langle T^{\mu \nu}(z) V^\alpha(x) V^\beta(0)\right\rangle 
\eeqa
according to
\beqa
\Gamma_D^{\mu\alpha\beta}(k,p)&=& 
i \frac{\partial}{\partial k^\nu}\Gamma^{\mu\nu\alpha\beta}(k,p) \,.
\eeqa
As we have already mentioned, this equation allows us to identify a pole term in the $J_DVV$ diagram from the corresponding pole 
structure in the $TVV$ vertex. 
In the following we will show the emergence of the anomaly poles in the QED and QCD cases. 

\subsection{The dilaton anomaly pole in the QED case}
\label{SecPoleQED}

For definiteness, it is convenient to briefly review the characterization of the $TVV$ vertex in the QED case with a massive fermion 
(see \cite{Armillis:2009pq} for more details).
The full amplitude  can be expanded in a specific basis of $13$ tensors first identitfied in \cite{Giannotti:2008cv}
\bea
\Gamma^{\mu\nu\alpha\beta}(p,q) =  \, \sum_{i=1}^{13} F_i (s; s_1, s_2,m^2)\phi_i^{\mu\nu\alpha\beta}(p,q)\,,
\label{Gamt}
\eea
where the $13$ invariant amplitudes $F_i$ are functions of the kinematical invariants $s=k^2=(p+q)^2$, \mbox{$s_1=p^2$}, $s_2=q^2$, 
with $p$ and $q$ the momenta of the external photons,  
and of the internal fermion mass $m$. The list of the tensor structures $\phi_i$  can be found in \cite{Giannotti:2008cv}. The number 
of these form factors reduces from $13$ to $3$ in the case of on-shell photons.  For our purposes, being interested in the appearance 
of the anomaly poles, we only need the contributions that generate a non zero trace. These come from the tensors 
$\phi_1^{\mu\nu\alpha\beta}$ and $\phi_2^{\mu\nu\alpha\beta}$ which are
\beqa	
\phi_1^{\mu\nu\alpha\beta}&=&\left(k^2 \eta^{\mu\nu} - k^{\mu } k^{\nu}\right) u^{\alpha\beta}(p,q), \nn \\
\phi_2^{\mu\nu\alpha\beta}&=&\left(k^2 \eta^{\mu\nu} - k^{\mu } k^{\nu}\right) w^{\alpha\beta}(p,q),
\eeqa
where
\bea
&&u^{\alpha\beta}(p,q) \equiv (p\cdot q) \eta^{\alpha\beta} - q^{\alpha}p^{\beta}\,,\nonumber \\
&&w^{\alpha\beta}(p,q) \equiv p^2 q^2 \eta^{\alpha\beta} + (p\cdot q) p^{\alpha}q^{\beta}
- q^2 p^{\alpha}p^{\beta} - p^2 q^{\alpha}q^{\beta}.\,
\label{uwdef}
\eea
For two on-shell final state photons ($s_1=s_2=0$) and a massive fermion we obtain
\beq
\label{oom}
{F_1 (s;\,0,\,0,\,m^2)} = 
F_{1\, pole} \,  + \, \frac{e^2 \,   m^2}{3 \, \pi ^2 \, s^2} \, - \frac{e^2 \, m^2}{3 \, \pi^2 \, s}  \, \mathcal C_0 (s, 0, 0, m^2,m^2,m^2) 
\bigg[\frac{1}{2}-\frac{2 \,m^2}{ s}\bigg],  \\
\eeq
where 
\beq F_{1\, pole}=- \frac{e^2 }{18 \, \pi^2 s} 
\eeq
and the scalar three-point function $ \mathcal C_0 (s, 0,0,m^2,m^2,m^2) $ is given by 
\beq
\mathcal C_0 (s, 0,0,m^2,m^2,m^2) = \frac{1}{2 s} \, \log^2 \frac{a_3+1}{a_3-1}, \qquad \mbox{with} \quad a_3 = \sqrt {1-4m^2/s} \,.
\eeq
In the massless fermion case two properties of this expansion are noteworthy: 1) the trace anomaly 
takes contribution only from a single tensor structure $(\phi_1)$ and invariant amplitude $(F_1)$ which coincides with the pole term; 
2) the residue of this pole as $s\to 0$ is nonzero, showing that the pole is coupled in the infrared. Notice that the form factor 
$F_2$, which in general gives a nonzero contribution to the trace in the presence of mass terms, is multiplied by a tensor 
structure ($\phi_2$) which {\em vanishes} when the two photons are on-shell.
Therefore, similarly to the case of the chiral anomaly, also in this case the anomaly is {\em entirely} given by the appearance 
of an anomaly pole. 
We stress that this result is found to be exact in dimensional regularization, which is a mass independent scheme:
at perturbative level, the anomalous breaking of the dilatation symmetry, related to an anomaly pole in the spectrum of all
the gauge-ivariant correlators studied in this work, is separated from the sources of {\em explicit} breaking. The latter are related to the mass parameters and/or to the gauge bosons virtualities $p^2$ and $q^2$. 

\begin{figure}[t]
\begin{center}
\includegraphics[scale=1.2]{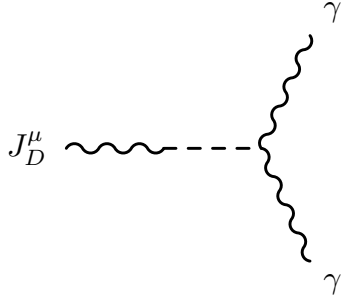}
\end{center}
\caption{Exchange of a dilaton pole mediated by the $J_D V V$ correlator. }
\label{dilatonpole}
\end{figure}

To analyze the implications of the pole behaviour discussed so far for the $TVV$ vertex and its connection with the $J_DVV$
correlator, we limit our attention on the anomalous contribution ($F_1\, \phi_1^{\mu\nu\alpha\beta}$), 
which we rewrite in the form
\beq
\Gamma_{pole}^{\mu\nu\alpha \beta}(k,p)\equiv
- \frac{e^2}{18\pi^2}\frac{1}{k^2}\left( \eta^{\mu \nu} k^2 - k^\mu k^\nu\right) u^{\alpha \beta}(p,q) \, , 
\qquad q = k - p \, .
\label{uref}
\eeq
This implies that the $J_D VV$ correlator acquires a pole as well
\beqa
\Gamma_{D\, pole}^{\mu\alpha\beta}
&=& 
- i \frac{e^2}{18 \pi^2}\frac{\partial}{\partial k^\nu}
\left[ \frac{1}{k^2}\, \left( \eta^{\mu\nu } k^2 - k^{\mu} k^{\nu}\right) u^{\alpha \beta}(p,k-p) \right]
\eeqa
and acting with the derivative on the right hand side we finally obtain 
\beq
\Gamma_{D \, pole}^{\mu\alpha\beta}(k,p)= 
i\, \frac{e^2}{6 \pi^2}\frac{k^\mu}{k^2}u^{\alpha \beta}(p,k-p) - i \frac{e^2}{18 \pi^2}\frac{1}{k^2}
\left( \eta^{\mu\nu } k^2 - k^{\mu} k^{\nu}\right)\frac{\partial}{\partial k_\nu}u^{\alpha \beta}(p,k-p).
\eeq
Notice that the first contribution on the right hand side of the previous equation corresponds to an anomaly pole, 
shown pictorially in Fig. \ref{dilatonpole}. In fact, by taking a derivative of 
the dilatation current only this term will contribute to the corresponding Ward identity
\beq
k_\mu \, \Gamma_D^{\mu\alpha\beta}(k,p) = i \frac{e^2}{6 \pi^2}u^{\alpha \beta}(p,k-p),
\label{res}
\eeq
which is the expression in momentum space of the usual relation $\partial J_D\sim FF$, while the second term trivially vanishes.  
Notice that the pole in (\ref{res}) has disappeared, and we are left just with its residue on the r.h.s., or, 
equivalently, the pole is removed in Eq. (\ref{uref}) if we trace the two indices $(\mu,\nu)$.

%

\subsection{The dilaton anomaly pole in the QCD case}

The analysis presented for the dilatation current of QED can be immediately generalized to the case of QCD.
Following a similar reasoning, we expand the on-shell $TVV$ vertex, with $V$ denoting now the gluon, as
\bea
\Gamma^{\mu\nu\alpha\beta}(p,q) = \delta^{a b} \sum_{i = 1}^{3} \Phi_i(s; 0,0) \, t_i^{\, \mu\nu\alpha\beta}(p,q) \qquad \mbox{with} 
\quad p^2 = q^2 = 0 \, ,
\eea
with the tensor basis given by
\bea
t_1^{\, \mu \nu \a \b} (p,q) &=&
(s \, \eta^{\mu\nu} - k^{\mu}k^{\nu}) \, u^{\a \b} (p,q),
\label{widetilde1} \nn \\
t_2^{\, \mu \nu \a \b} (p,q) &=& - 2 \, u^{\a \b} (p,q) \left[ s \, \eta^{\mu \nu} + 2 (p^\mu \, p^\nu + q^\mu \, q^\nu )
- 4 \, (p^\mu \, q^\nu + q^\mu \, p^\nu) \right],
\label{widetilde2} \nn \\
t^{\, \mu \nu \alpha \beta}_{3} (p,q) &=&
\big(p^{\mu} q^{\nu} + p^{\nu} q^{\mu}\big)g^{\alpha\beta}
+ \frac{s}{2} \left(\eta^{\alpha\nu} \eta^{\beta\mu} + \eta^{\alpha\mu} \eta^{\beta\nu}\right) \nn \\
&&
- \eta^{\mu\nu} \left(\frac{s}{2} \eta^{\alpha \beta}- q^{\alpha} p^{\beta}\right)
-\left(\eta^{\beta\nu} p^{\mu} + \eta^{\beta\mu} p^{\nu}\right)q^{\alpha}
- \big (\eta^{\alpha\nu} q^{\mu} + \eta^{\alpha\mu} q^{\nu }\big)p^{\beta},
\label{widetilde3}
\eea
where $\delta^{a b}$ is the diagonal matrix in color space. 
Again we have $s=k^2 = (p+q)^2$, with the virtualities of the two gluons being $p^2 = q^2 = 0$.
Notice that in the massless fermion limit only the first ($t_1$) of these 3 form factors contributes to the anomaly.
The corresponding on-shell form factors with massive quarks are
\bea
\Phi_{1}(s;0,0) &=& - \frac{g^2}{72 \pi^2 \,s}\left(2 n_f - 11 N_C \right) + \frac{g^2}{6 \pi^2}\sum_{i=1}^{n_f} m_i^2 \, \bigg\{ 
\frac{1}{s^2} \, - \, \frac{1} {2 s}\mathcal C_0 (s, 0, 0, m_i^2,m_i^2,m_i^2)
\bigg[1-\frac{4 m_i^2}{ s}\bigg] \bigg\}, \,
\label{Phi1} \nn \\
\Phi_{2}(s;0,0) &=& - \frac{g^2}{288 \pi^2 \,s}\left(n_f - N_C\right) \nn \\
&-& \frac{g^2}{24 \pi^2} \sum_{i=1}^{n_f} m_i^2 \, \bigg\{ \frac{1}{s^2}
+ \frac{ 3}{ s^2} \mathcal D_0 (s, 0, 0, m_i^2, m_i^2)
+ \frac{ 1}{s } \mathcal C_0(s, 0, 0, m_i^2,m_i^2,m_i^2 )\, \left[ 1 + \frac{2 m_i^2}{s}\right]\bigg\},
\label{Phi2} \nn \\
\Phi_{3}(s;0,0) &=& \frac{g^2}{288 \pi^2}\left(11 n_f - 65 N_C\right) - \frac{g^2 \, N_C}{8 \pi^2} 
\bigg[ \frac{11}{6} \mathcal B_0(s,0,0) - \mathcal B_0(0,0,0) +  s  \,\mathcal C_0(s,0,0,0,0,0) \bigg] \nn \\
&+&   \frac{g^2}{8 \pi^2} \sum_{i=1}^{n_f}\bigg\{  \frac{1}{3}\mathcal B_0(s, m_i^2,m_i^2) + m_i^2 \, \bigg[
\frac{1}{s}
 + \frac{5}{3 s}  \mathcal D_0 (s, 0, 0, m_i^2) + \mathcal C_0 (s, 0,0,m_i^2,m_i^2,m_i^2) \,\left[1 + \frac{2 m_i^2}{s}\right]
\bigg]\bigg\} , \nn \\ 
\label{Phi3}
\eea
where $m_i$ denotes the quark mass, and we have summed over the fermion flavours $(n_f)$, while $N_C$ denotes the number of colors.
Notice the appearance of the $1/s$ pole in $\Phi_1$, which saturates the contribution to the trace 
anomaly in the massless limit which becomes
\beq
\Phi_{1}(s;0,0) = - \frac{g^2}{72 \pi^2 \,s}\left(2 n_f - 11 N_C\right).
\label{polepole}
\eeq
As for the QED case, this is the only invariant amplitude which contributes to the anomalous trace part of the correlator.
The pole completely accounts for the trace anomaly and is clearly inherited by the QCD dilatation current, 
for the same reasonings discussed above.

\subsection{Mass corrections to the dilaton pole}

The discussion of the mass corrections to the massless dilaton can follow quite closely the strategy adopted in 
the pion case using partially conserved axial currents (PCAC) techniques. Also in this case, as for PCAC in the past, 
one can assume a partially conserved dilaton current (PCDC) in order to relate the decay amplitude of the dilaton $f_\rho$ 
to its mass $m_\rho$ and to the vacuum energy. \\
For this goal we define the 1-particle transition amplitudes for the dilatation current and the EMT between the vacuum and a dilaton 
state with momentum $p_\mu$
\beqa
\langle 0| J^\mu_D(x) |\rho, p \rangle 
&=& 
- i \, f_\rho \, p^\mu \, e^{-i p \cdot x} 
\nn \\
\langle 0| T^{\mu\nu}(x) |\rho, p \rangle 
&=&
\frac{f_\rho}{3} \, \left( p^\mu p^\nu - \eta^{\mu\nu} \, p^2\right)\, e^{-i p \cdot x},
\eeqa
both of them giving 
\beq
\label{rel1}
\partial_\mu \langle 0| J^\mu_D(x) |\rho, p\rangle = \eta_{\mu\nu} \langle 0| T^{\mu\nu}(x) |\rho, p \rangle = - f_\rho \, m_\rho^2 \, e^{-i p \cdot x}.
\eeq
We introduce the dilaton interpolating field $\rho(x)$ via a PCDC relation
\beq
\partial_\mu J^\mu_D(x) = - f_\rho \, m_\rho^2 \, \rho(x)
\label{pcdcrel}
\eeq
with 
\beq
\langle 0|\rho(x)|\rho, p\rangle = e^{-i p \cdot x}
\eeq
and the matrix element 
\beq
\mathcal A^\mu(q)= \int d^4 x \, e^{i q \cdot x} \,\langle 0| T \{ J^\mu_D(x) {T^\alpha}_\alpha(0) \} |0 \rangle,
\label{www}
\eeq
where $T\{\ldots\}$ denotes the time ordered product. \\
Using dilaton pole dominance we can rewrite the contraction of 	$q_\mu$ with this correlator as 
\bea
\label{inter}
\lim_{q_\mu \to 0} q_\mu \, \mathcal A^\mu(q)  = f_\rho \,\left\langle \rho, q=0| {T^\alpha}_\alpha(0) |0\right\rangle \,, 
\eea
where the soft limit $q_\mu \to 0$ with $q^2 \gg m_\rho^2 \sim 0$ has been taken. \\  
At the same time the dilatation Ward identity on the amplitude $\mathcal A^{\mu}(q)$ in Eq.(\ref{www}) gives 
\beqa
\label{WIJD}
q_\mu \mathcal A^\mu(q)
&=&
i \int d^4 x\,e^{i q \cdot x}\, \frac{\partial}{\partial x^\mu}\, 
\left\langle 0 \left| T\{ J^\mu_D(x) {T^\alpha}_\alpha(0) \} \right|0\right\rangle
\nn \\
&=&
i \int d^4 x \, e^{i q \cdot x} \,  
\left\langle 0 \left| T\{ \partial_\mu J^\mu_D(x) {T^\alpha}_\alpha(0)\} \right|0\right\rangle
+ i  \int d^4 x \, e^{i q \cdot x} \, \delta(x_0) \, 
\left\langle 0\left| \left[ J_D^0(x), {T^\alpha}_\alpha(0)\right] \right|0\right\rangle. 
\eeqa
The commutator of the time component of the dilatation charge density and the trace of the EMT can be rewritten as 
\beq
\left[ J_D^0(0, {\bf x}), {T^\alpha}_\alpha(0)\right] = -i \delta^3 ({\bf x})\left( 
d_T + x\cdot \partial\right){T^\alpha}_\alpha(0) 
\label{derr}
\eeq
where $d_T$ is the canonical dimension of the EMT $(d_T=4)$. 
Inserting Eq.(\ref{derr}) in the Ward identity (\ref{WIJD}) and neglecting the first term due to the nearly conserved dilatation current ($m_\rho \sim 0 $), 
we are left with 
\beq
q_\mu \mathcal A^\mu(q)= d_T \, \left\langle 0|{T^\alpha}_\alpha(0) |0\right\rangle.
\label{interdue}
\eeq
In the soft limit, with $q^2 \gg  m_\rho^2$, comparing Eq.(\ref{inter}) and Eq.(\ref{interdue}) we obtain 
\beqa
\lim_{q^\mu\to 0} q^\mu \mathcal{A}_\mu =
f_\rho \, \left\langle \rho, q=0| {T^\alpha}_\alpha(0) |0\right\rangle  = d_T \, \left\langle 0|{T^\alpha}_\alpha(0) |0\right\rangle.
\eeqa
Introducing the vacuum energy density $\epsilon_{vac}=\left\langle 0|T^0_0 |0\right\rangle = \frac{1}{4}  \left\langle 
0|T^\alpha_\alpha(0) |0\right\rangle$ and using the relation in Eq.(\ref{rel1}) we have 
\beq
\left\langle\rho, p=0| {T^\mu}_\mu |0\right\rangle =- f_\rho m_\rho^2 = \frac{d_T}{f_\rho} \epsilon_{vac}
\eeq
from which we finally obtain ($d_T = 4$)
\beq
f_\rho^2 m_\rho^2 = -16 \, \epsilon_{vac}. 
\eeq
This equation fixes the decay amplitude of the dilaton in terms of its mass and the vacuum energy. 
Notice that $\epsilon_{vac}$ can be related both to the anomaly and possibly to explicit contributions of the 
breaking of the dilatation symmetry since 
\beq
\label{beta}
\epsilon_{vac}= \frac{1}{4} \, \left\langle 0 \left| \frac{\beta(g)}{2 g} F_{\mu\nu}F^{\mu\nu} \right|0\right\rangle + ... 
\eeq
where the ellipsis saturate the anomaly equation with extra mass-dependent contributions, which may be far larger in size then the anomaly term. In (\ref{beta}) we have assumed, for simplicity, the coupling of the pole to a single gauge field, with 
a beta function $\beta(g)$, but obviously, it can be generalized to several gauge fields. 

 In the case of higher dimensional operators we would get 
\beq
\epsilon_{vac} = \frac{1}{4} \, \left\langle 0 \left| \frac{\beta(g)}{2 g} \, F_{\mu\nu}\,F^{\mu\nu} \right|0\right\rangle 
+ \sum_i g_i \, (d_i-4) \, \left\langle 0| O_i | 0 \right\rangle,
\eeq
valid around the scale at which the PCDC approximation holds. Therefore a massless pole can be corrected nonperturbatively according 
to some completion theory, causing its mass to shift. Perturbation theory 
gives indications about the interpolating fields which can couple to it, as we have seen by exploiting the chiral analogy, but not 
more than that. The corrections are model-dependent and can be the subject of additional 
phenomenological searches, but the dilatation current takes the role, with no doubt, of an interpolating field for the propagation of 
such a scalar intermediate state.

\section{The infrared coupling of an anomaly pole and the anomaly enhancement} 

It is easy to figure out from the results of the previous sections that the coupling of a (graviscalar) dilaton to the anomaly 
causes a large enhancement of its 2-photons and 2-gluons decays. One of the features of the graviscalar interaction is that its coupling includes anomalous contributions which are part both of the two-photon and of the two-gluon cross sections. For this reason, 
if an enhancement respect to the Standard Model rates is found only in one of these two channels and it is associated to the exchange of a spin zero intermediate state, this result could be used to rule out the exchange of a graviscalar.  

On the other hand, for an effective dilaton, identified by an anomaly pole in the $J_D VV$ correlator of the Standard Model, the case is more subtle, since the coupling of this effective state to the anomaly has to be introduced by hand. This state should be identified, in the perturbative picture,  with the corresponding anomaly pole. The situation, here, is closely similar to the pion case, 
 where the anomalous contribution to the pion-photon-photon vertex is added - a posteriori - to a Lagrangian which is otherwise chirally symmetric. 
Also in the pion case the anomaly enhancement can be justified, in a  perturbative approach, by the infrared coupling of the anomaly pole of the 
$AVV$ diagram.

  In general, in the case of an effective dilaton, one is allowed to write down a Lagrangian which is assumed to be scale invariant and, at a second 
  stage, introduce the direct coupling of this state to the trace anomaly. The possibility of coupling such a state to the photon and to the gluons or just to the photons, for instance, is a delicate issue for which a simple perturbative approach is unable to offer a definitive answer. 
For instance, if we insist that confinement does not allow us to have, in any case, on-shell final state gluons, the two-gluon coupling of an effective dilaton, identified in the corresponding $J_D VV$ correlator, should not be anomaly enhanced. In fact, with one or two off-shell final state gluons, the residue of the anomaly pole in this correlator is zero. Few more comments on this issue can be found in appendix \ref{AppDecoupling}.

We feel, however, that a simple perturbative analysis may not be completely sufficient to decide whether or not the coupling of such a state to the gluon anomaly takes place. On the other hand, there is no doubt, by the same reason, that such a coupling should occur in the 2-photon case, being the photons massless asymptotic states. In this case the corresponding anomaly pole of the $J_D \gamma\gamma$ vertex is infrared coupled.  

Similar enhancements are present in the case of quantum scale invariant extensions of the Standard Model \cite{Goldberger:2007zk},  where one assumes that the spectrum of the theory is extended with new massive states in order to set the $\beta$ functions of the gauge couplings to vanish. In a quantum scale invariant theory such as the one discussed in \cite{Goldberger:2007zk}, the dilaton couples only to massive states, but the heavy mass limit and the condition of the vanishing of the complete $\beta$ functions, leave at low energy a dilaton interaction proportional only to the $\beta$ functions of the low energy states. We have commented on this point in appendix \ref{quantum}. The "remnant" low energy interaction is mass-independent and coincides with that due to a typical anomalous coupling, although its origin is of different nature, since anomalous contributions are genuinely mass-independent. 

For this reason, the decays of a dilaton produced by such extensions carries "anomaly like" enhancements as in the graviscalar case. Obviously, such enhancements to the low energy states of the Standard Model would also be typical of the decay of a Higgs field, which couples proportionally to the mass of an intermediate state, if quantum scale invariance is combined with the decoupling of a heavy sector. This, in general, causes an enhancement of the Higgs decay rates into photons and gluons. A partial enhancement only of the di-photon channel could be accomplished, in this approach, by limiting the above quantum scale invariant arguments only to the electroweak sector.

\section{Conclusions}
We have presented a general discussion of dilaton interactions with the neutral currents sector of the Standard Model. In the case of a 
fundamental graviscalar as a dilaton, we have presented the complete electroweak corrections to the corresponding interactions and we have discussed the renormalization properties of the same vertices. In particular, we have shown that the renormalizability of the dilaton vertices is inherited directly from that of the Standard Model only if the Higgs sector is characterized by a conformal coupling ($\chi$) fixed at the value 1/6.

Then we have moved to an analysis of the analytic structure of the $J_D VV$ correlator, showing that it supports an anomaly pole as an interpolating state, which indicates that such a state can be interpreted as the Nambu-Goldstone (effective dilaton) mode of the anomalous breaking of the dilatation symmetry. 

In fact, the trace anomaly seems to bring in some important information concerning the dynamics of the Standard Model, aspects that we have 
tried to elucidate. For this reason, we have extended a previous analysis of ours of the $TVV$ vertex, performed in the broken electroweak phase and in QCD, in order to 
characterize the dynamical behaviour of the analogous $J_D VV$ correlator. The latter carries relevant information on the 
anomalous breaking of the dilatation symmetry in the Standard Model. In fact, as we move to high energy, far above the electroweak scale, the Lagrangian of the Standard Model 
becomes approximately scale-invariant. This approximate dilatation 
symmetry is broken by a quantum anomaly and its signature, as we have shown in our analysis, is in the appearance of 
an anomaly pole in the $J_DVV$ correlator. The same pole might appear in correlators with multiple insertions of $J_D$, but the proof 
of their existence is far more involved and requires further investigations. 
This pole is clearly massless in the perturbative picture, and accounts for the anomalous breaking 
of this approximate scale invariance. 

 \vspace{1cm}
 \centerline{\bf Acknowledgments} 
We thank Pietro Colangelo for discussions. This work is supported by INFN of Italy under Iniziativa Specifica BARI-21.

\appendix

\section{The coupling/decoupling of the anomaly pole}
\label{AppDecoupling}

In perturbative QCD, the anomalous $AVV$ diagram is characterized by the presence of an anomaly pole in the variable 
$k^2$, with $k$ denoting the momentum of the axial-vector current, which is explicitly shown in Eq.(\ref{AVVpole}). 
It is interesting to note that this structure has a non-vanishing residue for on-shell photons and for massless quarks running
in the loop. In this case the pole is said to be \emph{infrared coupled}.
This feature, supplemented by usual polology arguments, leads to a $\pi\to \gamma \gamma$ decay rate which 
is enhanced with respect to the non-anomalous case.
On the other hand, if the photons are virtual or the quarks are massive the anomaly pole decouples, namely, its residue is zero. 
We refer to \cite{Armillis:2009sm} for more details. The same behaviour is shared by the conformally anomalous $TVV$ diagram \cite{Armillis:2009pq}.
We illustrate this important point in the QED case by considering the off-shell correlator. 

We denote with $s_1$ and  $s_2$ the virtualities of the two final state photons and with $m$ the mass 
of the fermion running in the loops. 
The case with on-shell photons and a massive fermion has already been discussed in section \ref{SecPoleQED}. There we have shown that 
the anomaly pole has a non-vanishing residue only in the conformal limit, when all masses are set to zero. 
Indeed, in the case of a massive fermion, besides the fact that the anomaly pole anyway appears in the corresponding invariant 
amplitude $F_1 (s;\,0,\,0,\,m^2)$, as one can see from Eq.(\ref{oom}), it will decouple, showing a zero residue
\bea
\lim_{s\rightarrow0} \, s \, \Gamma^{\mu\nu\a\b}(s;\,0,\,0,\,m^2) =0 \,.
\eea
As for the chiral anomaly case, the absence of the internal fermion masses is not sufficient to guarantee the infrared coupling of 
the anomaly pole. Indeed, if $m=0$ but the photons are taken off-shell, being characterized by non zero virtualities $s_1$ and $s_2$, 
one can check that the entire correlator is completely free from anomaly poles as
\bea
\lim_{s\rightarrow0} \, s \, \Gamma^{\mu\nu\a\b}(s;\,s_1,\,s_2,\,0) =0 \,.
\eea
The computation of this limit needs the explicit results for all the invariant amplitudes $F_i$, which are not given here 
due to their lengthy expressions but can be found in \cite{Armillis:2009pq} where the coupling/decoupling features of the $TVV$ 
are discussed in more detail.

One should be aware of the fact that the same pole is present in the $AVV$ diagram when $VV$ are now the gluons. If the two gluons 
are on-shell, as in the 2-photon case, the perturbative anomaly pole is again infrared coupled. Obviously, such an an enhancement is 
not observable, since the gluons cannot be  on-shell, because of confinement. In the perturbative picture, a 
non-zero virtuality of the two gluons is then sufficient to exclude an infrared coupling of the anomaly pole.

\section{A classical scale invariant Lagrangian with a dilaton field}
\label{classical}

In this appendix we briefly describe the construction of a scale-invariant theory to clarify some of the issues
concerning the coupling of a dilaton. In particular, the example has the goal to 
illustrate that in a classical scale-invariant extension of a given theory, the dilaton couples only to operators which are mass dependent, 
and thus scale breaking, before the extension. We take the case of a fundamental dilaton field (not a composite) introduced 
in this type of extensions.

A scale invariant extension of a given Lagrangian can be obtained if we promote all the dimensionfull constants to dynamical 
fields. 
We illustrate this point in the case of a simple interacting scalar field theory incorporating the Higgs mechanism. 
At a second stage we will derive the structure of the dilaton interaction at order $1/\Lambda$, where 
$\Lambda$ is the scale characterizing the spontaneous breaking of the dilatation symmetry.

Our toy model consists in a real singlet scalar with a potential of the kind of $V_2(\phi)$ introduced in section \ref{NonGrav},
\beq
\label{original}
\mathcal L = \frac{1}{2} (\partial \phi)^2 - V_2(\phi) =
\frac{1}{2} (\partial \phi)^2 + \frac{\mu^2}{2}\, \phi^2 - \lambda\, \frac{\phi^4}{4} - \frac{\mu^4}{4\,\lambda}\, ,
\eeq
obeying the classical equation of motion
\beq \label{scalarEOM}
\square \phi = \mu^2\,\phi - \lambda\, \phi^3\, .
\eeq
Obviously this theory is not scale invariant due to the appearance of the mass term $\mu$. This feature is reflected in the trace of the EMT.
Indeed the canonical EMT of such a theory and its trace are
\bea
T^{\mu\nu}_{c}(\phi) 
&=& 
\partial^\mu \phi\, \partial^\nu \phi 
- \frac{1}{2}\,\eta^{\mu\nu} \bigg[ (\partial \phi)^2 + \mu^2 \,\phi^2 
-  \lambda\,\frac{\phi^4}{2} -  \frac{\mu^4}{2\,\lambda} \bigg] \, ,
\nn \\
{T_{c}^\mu}_\mu(\phi) &=& 
- (\partial\phi)^2 - 2\, \mu^2 \,\phi^2 +\lambda\, \phi^4 + \frac{\mu^4}{\lambda} \, .
\eea
It is well known that the EMT of a scalar field can be improved in such a way as to make its trace proportional only to the scale breaking parameter,
i.e. the mass $\mu$. This can be done by adding an extra contribution $T_I^{\mu\nu}(\phi, \chi)$ which is symmetric and conserved
\beq
T_I^{\mu\nu}(\phi,\chi)=\chi \left(\eta^{\mu\nu} \square \phi^2 - \partial^\mu \partial^\nu \phi^2 \right) \,,
\eeq
where the $\chi$ parameter is conveniently choosen.
The combination of the canonical plus the improvement EMT, 
$T^{\mu\nu} \equiv T_c^{\mu\nu} + T_I^{\mu\nu}$ has the off-shell trace
\beq
{T^\mu}_\mu(\phi,\chi)= 
(\partial\phi)^2\, \left( 6 \chi - 1 \right) - 2\, \mu^2\, \phi^2 
+ \lambda\, \phi^4 + \frac{\mu^4}{\lambda} + 6 \chi \phi\, \square \phi\, .
\eeq
Using the equation of motion (\ref{scalarEOM}) and chosing $\chi=1/6$ the trace relation given above 
becomes proportional uniquely to the scale breaking term $\mu$  
\beq \label{ImprovedTrace}
{T^\mu}_\mu(\phi,1/6) = - \mu^2 \phi^2 + \frac{\mu^4}{\lambda} \, .
\eeq
The scale invariant extension of the Lagrangian given in Eq.(\ref{original}) is achieved by promoting the mass terms to dynamical fields by the replacement 
\beq
\label{rep}
\mu \to \frac{\mu}{\Lambda} \, \Sigma,
\eeq
obtaining
\beq
\label{sigmaphi}
\mathcal L = 
\frac{1}{2}\, (\partial \phi)^2 +\frac{1}{2} (\partial \Sigma)^2 
+  \frac{ \mu^2}{2\,\Lambda^2}\, \Sigma^2\, \phi^2 - \lambda \frac{\phi^4}{4}
-  \frac{\mu^4}{4\,\lambda \, \Lambda^4}\, \Sigma^4
\eeq
where we have used Eq.(\ref{rep}) and introduced a kinetic term for the dilaton $\Sigma$. 
Obviously, the new Lagrangian is dilatation invariant, as one can see from the trace of the improved EMT 
\beq
{T^{\mu}}_{\mu}(\phi,\Sigma,\chi,\chi') = \left( 6\, \chi - 1 \right)\, (\partial\phi)^2
+ \left( 6 \chi^\prime -1\right)\, (\partial\Sigma)^2 
+ 6 \chi\, \phi\, \square \phi + 6 \chi^\prime\, \Sigma\, \square \Sigma 
- 2\, \frac{\mu^2}{\Lambda^2}\, \Sigma^2\, \phi^2 + \lambda\, \phi^4
+ \frac{1}{\lambda}\,\frac{\mu^4}{\Lambda^4}\,\Sigma^4 \,,
\eeq
which vanishes upon using the equations of motion for the $\Sigma$ and $\phi$ fields,
\bea
\square \phi &=& \frac{\mu^2}{\Lambda^2}\, \Sigma^2\, \phi -  \lambda\, \phi^3\, ,
\nn \\
\square \Sigma &=& \frac{\mu^2}{\Lambda^2}\, \Sigma\, \phi^2 - \frac{1}{\lambda}\, \frac{\mu^4}{\Lambda^4}\,\Sigma^3  \, ,
\eea
and setting the $\chi, \chi'$ parameters at the special value $\chi=\chi^\prime=1/6$. 

As we have already discussed in section \ref{NonGrav}, the scalar potential $V_2$ allows to perform 
the spontaneous breaking of the scale symmetry around a stable minimum point, 
giving the dilaton and the scalar field the vacuum expectation values $\Lambda$ and $v$ respectively
\beq
\Sigma  =  \Lambda + \rho \, , \quad \phi = v + h\, .
\eeq
For our present purposes, it is enough to expand the Lagrangian (\ref{sigmaphi}) around the vev for the dilaton field, 
as we are interested in the structure of the couplings of its fluctuation $\rho$
\beq \label{Manifest}
\mathcal L = \frac{1}{2}\, (\partial\phi)^2 + \frac{1}{2}\, (\partial\rho)^2 + \frac{\mu^2}{2}\, \phi^2 
- \lambda \, \frac{\phi^4}{4} - \frac{\mu^4}{4\,\lambda}
- \frac{\rho}{\Lambda}\,\left(- \mu^2\, \phi^2  + \frac{\mu^4}{\lambda}\right) + \dots\, ,
\eeq
where the ellipsis refer to terms that are higher order in $1/\Lambda$.
It is clear, from (\ref{ImprovedTrace}) and (\ref{Manifest}), that one can write an dilaton Lagrangian
at order $1/\Lambda$, as
\beq \label{RhoInteraction}
\mathcal L_{\rho} = (\partial\rho)^2 - \frac{\rho}{\Lambda}\, {T^{\mu}}_{\mu}(\phi,1/6) + \dots\, ,
\eeq
where the equations of motion have been used in the trace of the energy momentum tensor.
Expanding the scalar field around $v$ would render the previous equation more complicated and we omit it for definiteness.
We only have to mention that a mixing term $\sim \rho\, h$ shows up and it has to be removed diagonalizing 
the mass matrix, switching from interaction to mass eigenstates exactly in the way we discussed in section \ref{NonGrav}, 
to which we refer for the details. 

It is clear, from this simple analysis, that a dilaton, in general, does not couple to the anomaly,
but only to the sources of explicit breaking of scale invariance, i.e. to the mass terms.
The coupling of a dilaton to an anomaly is, on the other hand, necessary, 
if the state is interpreted as a composite pseudo Nambu-Goldstone mode of the dilatation symmetry.
Thus, this coupling has to be introduced by hand, in strict analogy with the chiral case.

\section{Quantum conformal invariance and dilaton couplings at low energy}
\label{quantum}

As a second example, we consider the situation in which all the SM fields are embedded in a (quantum) Conformal Field Theory (CFT) extension \cite{Goldberger:2007zk}
and we discuss the (loop-induced) couplings of the dilaton to the massless gauge bosons. 
At tree level the dilaton of \cite{Goldberger:2007zk} couples to the SM fields only through their masses, as the fundamental dilaton which we have discussed previously, and, in this respect, it behaves like the SM Higgs, without scale anomaly contributions. 
For this reason the dilaton interaction with the massless gauge bosons is induced by quantum effects mediated by heavy particles
running in the loops (in this context heavier or lighter is referred to the dilaton mass), and not by anomalous terms.
When the mass $m_i$ of the particle running in the loop is much greater than the dilaton mass, the coupling to the massless 
gauge bosons becomes 
\bea
\mathcal L_{\rho} = \frac{\alpha_s}{8 \pi} \sum_i b_g^i \, \frac{\rho}{\Lambda} (F_{g \, \mu\nu}^a)^2 Ê 
+ Ê \frac{\alpha_{em}}{8 \pi} \sum_i b_{em}^i \, \frac{\rho}{\Lambda} (F_{\gamma \, \mu\nu})^2 Ê \,,
\eea
where $b_{em}^i$ and $b_{g}^i$ are the contributions of the heavy field $i$ to the one-loop $\beta$ function (computed in the 
$\overline{MS}$ scheme) for the electromagnetic and strong coupling constants respectively. The $\beta$ functions are normalized as
\bea
\beta_i = \frac{g^3}{16 \pi^2} b^i \,.
\eea
Note that this result is independent from the heavy mass $m_i$ as one can prove by analyzing the structure of the mass corrections 
of the dilaton coupling, which reads as
\beq
\Gamma_{\rho V V}\sim \frac{g^2}{\pi^2 \Lambda} \, m_i^2 \, 
\bigg[ \frac{1}{s}  -  \frac{1} {2 }\mathcal C_0 (s, 0, 0, m_i^2, m_i^2, m_i^2) \bigg(1-\frac{4 m_i^2}{ s}\bigg)\bigg] 
\sim 
\frac{g^2}{\pi^2 \Lambda} \, \frac{1}{6} + O\left( \frac{s}{m_i^2} \right)
\eeq
where $s=m_{\rho}^2$ is fixed at the dilaton mass and we have performed the large mass limit of the amplitude using 
\beqa
\mathcal C_0(s, 0, 0, m_i^2, m_i^2, m_i^2) \sim - \frac{1}{2 m_i^2} \left( 1 + \frac{1}{12} \frac{s}{m_i^2} + O(\frac{s^2}{m_i^4} ) \right)
\eeqa
valid for $m_i^2 \gg s = m_{\rho}^2$. This shows that in the case of heavy fermions, 
the dependence on the fermion mass cancels. 
Obviously, this limit generates an effective coupling which is proportional to the $\beta$ function related to the heavy flavours. 
The same reasonings can be employed to the Higgs case as well. It clear that this coupling to the massless gauge bosons is dependent 
from new heavy states and, therefore, from the UV completion of the SM. This is certainly the case for the Standard Model Higgs whose 
double photon decay is one of the most important decay channel for new physics discoveries. \\ 
For the dilaton case the situation is slightly different. Surely we do not understand the details of the CFT extension, nor its 
particle spectrum, but nevertheless we know that the conformal symmetry is realized at the quantum level. Therefore the complete 
$\beta$ functions, including the contribution from all states, must vanish
\bea
\beta = \frac{g^3}{16 \pi^2} Ê\bigg[ \sum_{i} b^i Ê+ \sum_{j} b^j \bigg] = 0 \,,
\eea
where $i$ and $j$ run over the heavy and light states respectively. Exploiting the consequence of the quantum conformal symmetry, the 
dilaton couplings to the massless gauge bosons become
\bea
\mathcal L_{\rho} = 
- \frac{\alpha_s}{8 \pi} \sum_j b_g^j \, \frac{\rho}{\Lambda} (F_{g \, \mu\nu}^a)^2 Ê - Ê \frac{\alpha_{em}}{8 \pi}
\sum_j b_{em}^j \, \frac{\rho}{\Lambda} (F_{\gamma \, \mu\nu})^2 Ê \,,
\eea
in which the dependence from the $\beta$ functions of the light states is now explicit. We emphasize that the appearance of the light 
states contributions to the $\beta$ functions is a consequence of the vanishing of the complete $\beta$, and, therefore, of the CFT 
extension and not the result of a direct coupling of the dilaton to the anomaly.

\section{Appendix. Feynman rules}
\label{rules}
The Feynman rules used throughout the paper are collected here. We have 
\begin{itemize}
%
\item{ dilaton - gauge boson - gauge boson vertex}
\\ \\
\begin{minipage}{95pt}
\includegraphics[scale=1.0]{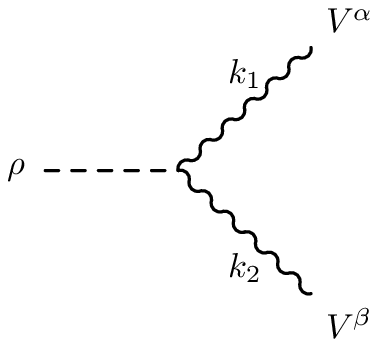}
\end{minipage}
\begin{minipage}{70pt}
\beqa
= V^{\alpha\beta}_{\rho VV}(k_1,k_2) =
- \frac{2\, i}{\Lambda} \bigg\{M_V^2 \, \eta^{\alpha\beta} 
- \frac{1}{\xi}\, \left( k_1^\alpha\, k_1^\beta + k_2^\alpha\, k_2^\beta + 2\, k_1^\alpha\, k_2^\beta \right) \bigg\}
\nn
\eeqa
\end{minipage}
\beqa
\label{FRdilVV}
\eeqa
where $V$ stands for the gluons or for the vector gauge bosons $A, Z$ and $W^{\pm}$ and,
if the gauge bosons are gluons, a color-conserving $\delta_{ab}$ matrix must be included.
\\ \\
\item{dilaton - fermion - fermion vertex}
\\ \\
\begin{minipage}{95pt}
\includegraphics[scale=1.0]{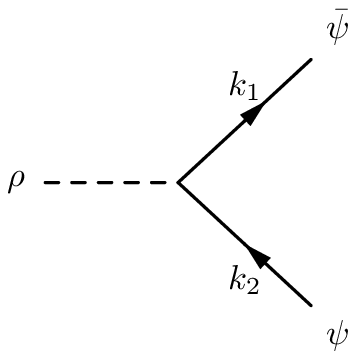}
\end{minipage}
\begin{minipage}{70pt}
\beqa
&=& 
V_{\rho \bar\psi\psi}(k_1,k_2) = 
- \frac{i}{2\,\Lambda} \, \bigg\{ 3 \, \left( \ksl_1 - \ksl_2 \right) + 8 \, m_f \bigg\}
\nn
\eeqa
\end{minipage}
\beqa
\label{FRdilFF}
\eeqa
If the fermions are quarks, the vertex must be multiplied by the identity color matrix $\delta_{a b}$. \\ \\
\item{dilaton - ghost - ghost vertex }
\\ \\
\begin{minipage}{95pt}
\includegraphics[scale=1.0]{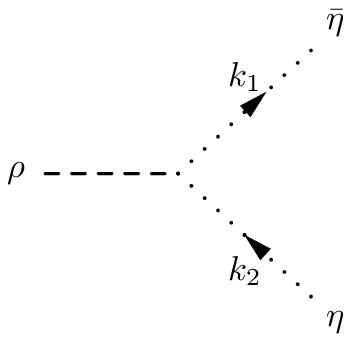}
\end{minipage}
\begin{minipage}{70pt}
\beqa
&=& 
V_{\rho \bar c c}(k_1,k_2)
= - \frac{2 \, i}{\Lambda} \bigg\{ k_1 \cdot k_2 + 2 \, M_{\eta}^2\bigg\}
\nn
\eeqa
\end{minipage}
\beqa
\label{FRdilUU}
\eeqa
where $\eta$ denotes both the QCD ghost fields $c^{a}$ or the electroweak ghost fields $\eta^{+}$, $\eta^{-}$ ed $\eta^Z$. In the QCD 
case one must include a color-conserving $\delta_{ab}$ matrix. 
\\ \\
\item{dilaton - scalar - scalar vertex}
\\ \\
\begin{minipage}{95pt}
\includegraphics[scale=1.0]{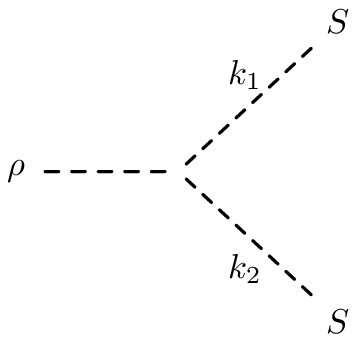}
\end{minipage}
\begin{minipage}{70pt}
\beqa
=  
V_{\rho SS}(k_1,k_2)
&=& - \frac{2 \, i}{\Lambda} \bigg\{ k_1 \cdot k_2 + 2 \, M_S^2  \bigg\} 
\nn \\
&=&  
\frac{i}{\Lambda} \,6\,\chi\, (k_1+k_2)^2 \nn
\eeqa
\end{minipage}
\beqa
\label{FRdilSS}
\eeqa
where $S$ stands for the Higgs $H$ and the Goldstones $\phi$ and  $\phi^{\pm}$. 
The first expression is the contribution coming from 
the minimal energy-momentum tensor while the second is due to the term of improvement.
\\ \\
\item{dilaton - Higgs vertex}
\\ \\
\begin{minipage}{95pt}
\includegraphics[scale=1.0]{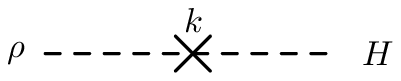}
\end{minipage}
\begin{minipage}{70pt}
\beqa
\qquad 
&=& 
V_{I,\, \rho H }(k)
= - \frac{i}{\Lambda} \frac{12\,\chi\, s_w M_W}{e} \, k^2 \nn
\eeqa
\end{minipage}
\beqa
\label{FRdilH}
\eeqa
This vertex is derived from the term of improvement of the energy-momentum tensor and it is a feature of the electroweak symmetry 
breaking because it is proportional to the Higgs vev.
\\ \\
\item{ dilaton - three gauge boson vertex}
\\ \\
\begin{minipage}{95pt}
\includegraphics[scale=1.0]{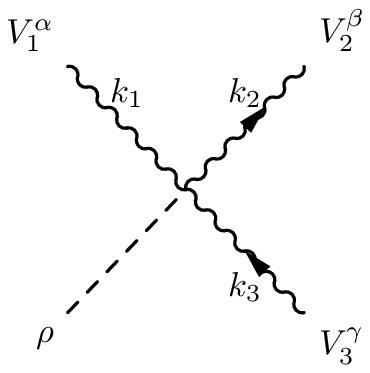}
\end{minipage}
\begin{minipage}{70pt}
\beqa
&=&  
V^{\alpha\beta\gamma}_{\rho V V V}
= 0
\nn
\eeqa
\end{minipage}
\beqa
\label{FRdilVVV}
\eeqa
where $V_1$, $V_2$, $V_3$ stand for gluon, photon, $Z$ and $W^{\pm}$.
\\ \\
\item{dilaton - gauge boson - scalar - scalar vertex }
\\ \\
\begin{minipage}{95pt}
\includegraphics[scale=1.0]{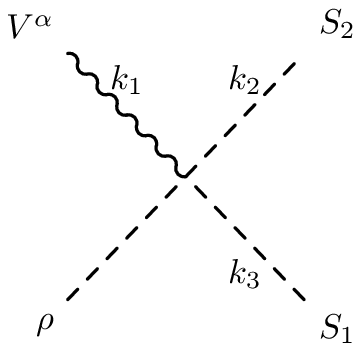}
\end{minipage}
\begin{minipage}{70pt}
\beqa
&=&
V^\alpha_{\rho VSS}(k_2,k_3)
=  - \frac{2\,i}{\Lambda}\, e \, \mathcal C_{V S_1 S_2} \, \left( k_2^\alpha - k_3^\alpha \right)
\nn
\eeqa
\end{minipage}
\beqa
\label{FRdilVSS}
\eeqa
with $\mathcal C_{V S_1 S_2}$ given by
\beqa
\mathcal C_{A\phi^{+}\phi^{-}} = 1 \qquad
\mathcal C_{Z\phi^{+}\phi^{-}} = \frac{c_w^2 - s_w^2}{2 s_w \, c_w} \qquad
\mathcal C_{Z H \phi} =  \frac{i}{2 s_w \, c_w}. \nn
\eeqa
\\ \\
\item{dilaton - gauge boson - ghost - ghost vertex}
\\ \\
\begin{minipage}{95pt}
\includegraphics[scale=1.0]{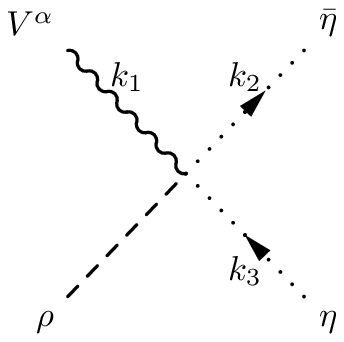}
\end{minipage}
\begin{minipage}{70pt}
\beqa
&=&
V^\alpha_{\rho V \bar\eta\eta}(k_2)
= - \frac{2\,i}{\Lambda} \, \mathcal C_{V \eta} \, k_2^\alpha
\nn
\eeqa
\end{minipage}
\beqa
\label{FRdilVUU}
\eeqa
where $V$ denotes the $g^a$, $A$, $Z$ gauge bosons and $\eta$ the ghosts $c^b$, $\eta^{+}$, $\eta^{-}$.
The coefficients $\mathcal C$ are defined as
\beqa
\mathcal C_{g^a  c^b} = f^{abc}\, g \qquad
\mathcal C_{A \eta^{+}} = e \qquad
\mathcal C_{A \eta^{-}} = -e \qquad
\mathcal C_{Z \eta^{+}} =  e \, \frac{c_w}{s_w} \qquad
\mathcal C_{Z \eta^{-}} =  - e \, \frac{c_w}{s_w}. \nn
\eeqa
%

\item{dilaton - gauge boson - gauge boson - scalar vertex}
\\ \\
\begin{minipage}{95pt}
\includegraphics[scale=1.0]{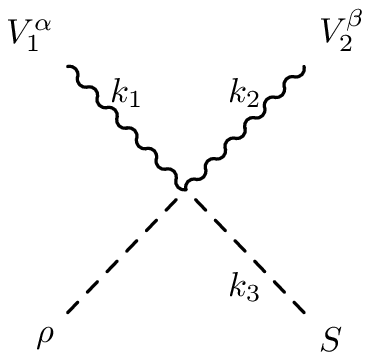}
\end{minipage}
\begin{minipage}{70pt}
\beqa
&=&
V^{\alpha\beta}_{\rho VVS}
= - \frac{2}{\Lambda} \, e \, \mathcal C_{V_1 V_2 S} \, M_W \, \eta^{\alpha\beta}
\nn
\eeqa
\end{minipage}
\beqa
\label{FRdilVVS}
\eeqa
where $V$ stands for $A$, $Z$ and $W^{\pm}$ and $S$ for $\phi^{\pm}$
and $H$. The coefficients are defined as
\beqa
\mathcal C_{A W^{+} \phi^{-}} = 1 \qquad
\mathcal C_{A W^{-} \phi^{+}} = -1 \qquad
\mathcal C_{Z W^{+} \phi^{-}} = - \frac{s_w}{c_w} \qquad \nn \\
\mathcal C_{Z W^{-} \phi^{+}} = \frac{s_w}{c_w} \qquad
\mathcal C_{Z Z H} = - \frac{i}{s_w \, c_w^2} \qquad
\mathcal C_{W^{+} W^{-} H} = - \frac{i}{c_w}. \nn
\eeqa
\\ \\
\item{dilaton - scalar - ghost - ghost vertex}
\\ \\
\begin{minipage}{95pt}
\includegraphics[scale=1.0]{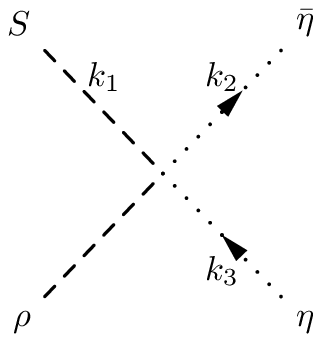}
\end{minipage}
\begin{minipage}{70pt}
\beqa
&=&
V{\rho S\bar\eta\eta}
= - \frac{4 \, i}{\Lambda}\, e \, \mathcal C_{S \eta} \, M_W
\nn
\eeqa
\end{minipage}
\beqa
\label{FRdilSUU}
\eeqa
where $S = H$ and $\eta$ denotes $\eta^{+}$, $\eta^{-}$ and $\eta^{z}$. The vertex is defined with the coefficients
\beqa
\mathcal C_{H \eta^{+}} = \mathcal C_{H \eta^{-}} = \frac{1}{2 s_w} \qquad \mathcal C_{H \eta^{z}} = \frac{1}{2 s_w \, c_w}. 
\nn
\eeqa
%

\item{dilaton - three scalar vertex}
\\ \\
\begin{minipage}{95pt}
\includegraphics[scale=1.0]{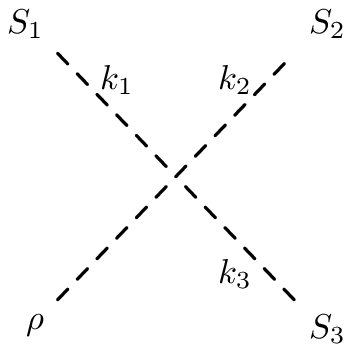}
\end{minipage}
\begin{minipage}{70pt}
\beqa
&=&
V_{\rho SSS}
= - \frac{4 \, i}{\Lambda} \, e \, \mathcal C_{S_1 S_2 S_3}
\nn
\eeqa
\end{minipage}
\beqa
\label{FRdilSSS}
\eeqa
with $S$ denoting $H$, $\phi$ and $\phi^{\pm}$. We have defined the coefficients
\beqa
\mathcal C_{H \phi \phi} = \mathcal C_{H \phi^{+} \phi^{-}} = \frac{1}{2 s_w \, c_w}
\frac{M_H^2}{M_Z} \qquad \mathcal C_{H H H} = \frac{3}{2 s_w \, c_w} \frac{M_H^2}{M_Z}. 
\nn
\eeqa
\\ \\
\item{dilaton - scalar - fermion - fermion vertex}
\\ \\
\begin{minipage}{95pt}
\includegraphics[scale=1.0]{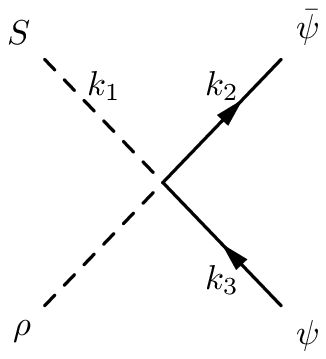}
\end{minipage}
\begin{minipage}{70pt}
\beqa
&=&
V_{\rho S\bar\psi\psi}
= - \frac{2\,i}{\Lambda} \, \frac{e}{s_w \, c_w} \frac{m_f}{M_Z}
\nn
\eeqa
\end{minipage}
\beqa
\label{FRdilSFF}
\eeqa
where  $S$ is only the Higgs scalar $H$.
\\ \\
\item{dilaton - gluon - fermion - fermion vertex}
\\ \\
\begin{minipage}{95pt}
\includegraphics[scale=1.0]{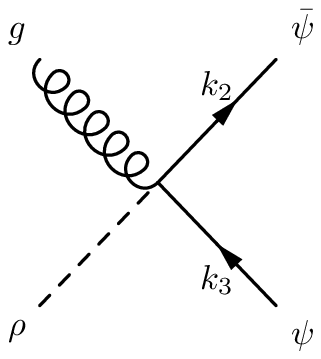}
\end{minipage}
\begin{minipage}{70pt}
\beqa
&=&
V^{a\,\alpha}_{\rho g\bar\psi\psi}
= \frac{3\, i}{\Lambda}\, g \, T^a \, \gamma^\alpha\, .
\nn
\eeqa
\end{minipage}
\beqa
\label{FRdilgFF}
\eeqa
\\ \\
\item{dilaton - photon - fermion - fermion vertex}
\\ \\
\begin{minipage}{95pt}
\includegraphics[scale=1.0]{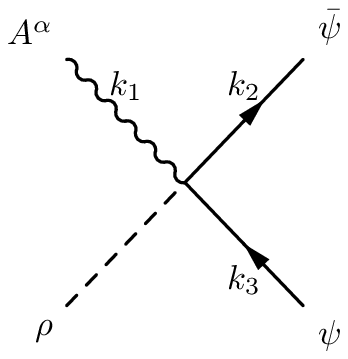}
\end{minipage}
\begin{minipage}{70pt}
\beqa
&=&
V^{\alpha}_{\rho \gamma\bar\psi\psi}
= Q_f \, e \frac{3\,i}{\Lambda}\, \gamma^\alpha
\nn
\eeqa
\end{minipage}
\beqa
\label{FRdilAFF}
\eeqa
where $Q_f$ is the fermion charge expressed in units of $e$.
\\ \\
\item{dilaton - Z - fermion - fermion vertex}
\\ \\
\begin{minipage}{95pt}
\includegraphics[scale=1.0]{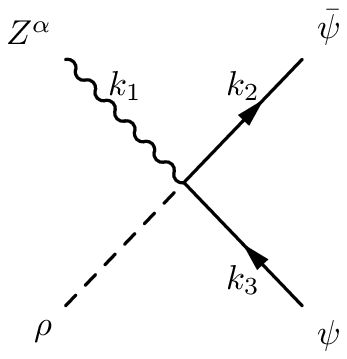}
\end{minipage}
\begin{minipage}{70pt}
\beqa
&=&
V^{\alpha}_{\rho Z\bar\psi\psi}
= \frac{3 \,i }{2\, \Lambda s_w \, c_w} \, e \, (C_v^f - C_a^f \gamma^5) \, \gamma^\alpha
\nn
\eeqa
\end{minipage}
\beqa
\label{FRdilZFF}
\eeqa
where $C_v^f$ and $C_a^f$ are the vector and axial-vector couplings of the $Z$ gauge boson
to the fermion ($f$). Their expressions are
\beqa
C_v^f = I^f_3 - 2 s_w^2 \, Q^f \qquad \qquad C_a^f = I^f_3. \nn
\eeqa
$I^f_3$ denotes the 3rd component of the isospin.
\\ \\
\item{dilaton - four gauge bosons vertex}
\\ \\
\begin{minipage}{95pt}
\includegraphics[scale=1.0]{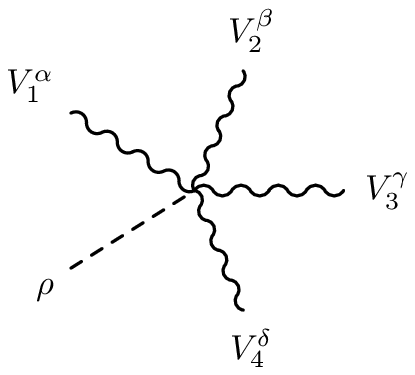}
\end{minipage}
\begin{minipage}{70pt}
\beqa
&=&
V^{\alpha\beta\gamma\delta}_{\rho VVVV}
\qquad \qquad = 0 \nn
\eeqa
\end{minipage}
\beqa
\label{FRdilVVVV}
\eeqa
where $V_1$, $V_2$, $V_3$ and $V_4$ denote $g$, $A$, $Z$ or $W^{\pm}$.
\\ \\
\item{dilaton - gauge boson - gauge boson - scalar - scalar vertex}
\\ \\
\begin{minipage}{95pt}
\includegraphics[scale=1.0]{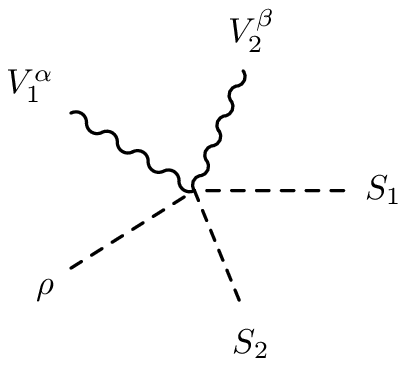}
\end{minipage}
\begin{minipage}{70pt}
\beqa
&=&
V^{\alpha\beta}_{\rho VVSS}
\qquad = \frac{2 \, i}{\Lambda} \, e^2 \, \mathcal C_{V_1 V_2 S_1 S_2} \, \eta^{\alpha\beta}
\nn
\eeqa
\end{minipage}
\beqa
\label{FRdilVVSS}
\eeqa
where $V_1$ and $V_2$ denote the neutral gauge bosons $A$ and $Z$, while the possible scalars are
$\phi$, $\phi^{\pm}$ and $H$. The coefficients are
\beqa
\mathcal C_{A A \phi^+ \phi^-} = 2
\qquad
\mathcal C_{A Z \phi^+ \phi^-} = \frac{c_w^2 - s_w^2}{s_w \, c_w}\qquad \mathcal C_{Z Z \phi^+ \phi^-} = 
\frac{\left( c_w^2 - s_w^2 \right)^2}{2 s_w^2 \,c_w^2}
\qquad
\mathcal C_{Z Z \phi \phi} =  \mathcal C_{Z Z H H} = \frac{1}{2 s_w^2 \, c_w^2} . \nn
\eeqa
\end{itemize}

\section{Appendix. The scalar integrals}
\label{scalars}

We collect in this appendix the definition of the scalar integrals appearing in the computation of the correlators.
One-, two- and three-point functions are denoted, respectively as $\mathcal A_0$, $\mathcal B_0$ and $\mathcal C_0$, with
\bea
\mathcal A_0 (m_0^2) &=& \frac{1}{i \pi^2}\int d^n l \, \frac{1}{l^2 - m_0^2} \,, \nn\\
\mathcal B_0 (k^2, m_0^2,m_1^2) &=&  \frac{1}{i \pi^2} \int d^n l \, \frac{1}{(l^2 - m_0^2) \, ((l + k )^2 - m_1^2 )} \,, \nn \\
\mathcal C_0 ((p+q)^2, p^2, q^2, m_0^2,m_1^2,m_2^2) &=& \frac{1}{i \pi^2} \int d^n l \, \frac{1}{(l^2 - m_0^2) \, ((l + p )^2 - m_1^2 ) \, ((l -q )^2 - m_2^2 ) } \,.
\eea
We have also used the finite combination of two-point scalar integrals
\bea
\mathcal D_0 (p^2, q^2, m_0^2,m_1^2) = \mathcal B_0 (p^2, m_0^2,m_1^2) - \mathcal B_0 (q^2, m_0^2,m_1^2) \,.
\eea
The explicit expressions of $\mathcal A_0$, $\mathcal B_0$ and $\mathcal C_0$ can be found in \cite{Denner:1991kt}.

\section{Appendix. Contributions to $\mathcal V_{\rho Z Z }$}
\label{VZZ}

\begin{itemize}
\item {\bf The fermion sector}
%
\beqa
&&
\Phi^{F}_{ZZ}(p,q)
=\sum_f \bigg\{
\frac{ \alpha  \, m_f^2}{\pi\,  s c_w^2 s_w^2 \left(s-4 M_Z^2\right)} \left(s-2 M_Z^2\right)
\left(C_a^{f \, 2} + C_v^{f\, 2}\right)
\nn \\
&&
+ \frac{2\, \alpha  \, m_f^2}{ \pi\, s c_w^2 \left(s-4 M_Z^2\right)^2 s_w^2}
\left[\left(2 M_Z^4-3 s M_Z^2+s^2\right) C_a^{f \, 2}+C_v^{f \, 2} M_Z^2 \left(2 M_Z^2 + s \right)\right]
\mathcal D_0(s, M_Z^2,m_f^2,m_f^2)
\nn \\
&&
+ \frac{\alpha \, m_f^2}{2 \,\pi \, s c_w^2 \left(s - 4 M_Z^2\right){}^2 s_w^2}
\left(s-2 M_Z^2\right) \big\{\left[4 M_Z^4-2 \left(8 m_f^2+s \right) M_Z^2+s \left(4 m_f^2+s \right)\right] C_a^{f\,2}
\nn \\
&&
+ C_v^{f \, 2} \left[4 M_Z^4+2 \left(3 s - 8 m_f^2\right) M_Z^2-s \left(s - 4 m_f^2\right) \right] \big\}
\vphantom{\frac{m_f^2}{\left(s-4 M_Z^2\right)}}\, \mathcal C_0(s, M_Z^2,M_Z^2 ,m_f^2,m_f^2,m_f^2)\bigg\}
\\
&&
\Xi^{F}_{ZZ}(p,q)
=\sum_f \bigg\{
- \frac{\alpha\, m_f^2}{\pi \,  s c_w^2 \left(s-4 M_Z^2\right) s_w^2} 
\left[\left(2 M_Z^4-4 s M_Z^2+s^2\right) C_a^{f \, 2}+2 C_v^{f \, 2} M_Z^4 \right]
\nn \\
&&
- \frac{\alpha\,m_f^2}{ \pi\, c_w^2 s_w^2}C_a^{f\,2}\mathcal B_0(s,m_f^2,m_f^2)
- \frac{2\,\alpha\,  m_f^2 \,M_Z^2}{ \pi \, s c_w^2 \left(s-4 M_Z^2\right){}^2 s_w^2} 
\big[s^2 C_a^{f \, 2} - 2 \left( C_a^{f \, 2} + C_v^{f \, 2} \right) M_Z^4
\nn \\
&&
+ 2 s \left(C_v^{f \, 2}-C_a^{f \, 2}\right) M_Z^2 \big]\, \mathcal D_0(s, M_Z^2, m_f^2,m_f^2) 
\nn \\
&&
- \frac{\alpha \, m_f^2}{ \pi\, s c_w^2 \left(s - 4 M_Z^2\right){}^2 s_w^2}
\bigg[ [ 4 M_Z^8-2 \left(8 m_f^2 + 5 s \right) M_Z^6 + 3 s \left(12 m_f^2+s \right) M_Z^4 - 16 s^2 m_f^2 M_Z^2 
\nn \\
&&
+ 2 s^3 m_f^2 ] C_a^{f \,2} + C_v^{f \, 2} M_Z^4 \left[4 M_Z^4-2 \left(8 m_f^2+s \right) M_Z^2 
+ s \left(4 m_f^2+s \right)\right] \bigg]\, 
\mathcal C_0(s, M_Z^2,M_Z^2 ,m_f^2,m_f^2,m_f^2) \vphantom{\frac{m_f^2}{\left(s-4 M_Z^2\right)}} \bigg\}
\nn \\
\eeqa

\item{\bf The $W$ boson sector}
%
\beqa
&&
\Phi^{W}_{ZZ}(p,q)
=
\frac{\alpha}{ s_w^2 \, c_w^2 \, \pi }
\bigg[ 
\frac{M_Z^2}{2\,s\left(s-4 M_Z^2\right)} \big[ 2 M_Z^2 \left(-12 s_w^6+32 s_w^4-29 s_w^2+9\right) 
\nn\\
&&
+ s \left(12 s_w^6-36 s_w^4+33 s_w^2-10\right)\big] \bigg]  
\nn \\
&&
+ \frac{\alpha \, M_Z^2}{ 2\, s_w^2 \, c_w^2 \,\pi\,s\,(s-4 M_Z)^2} 
[ 4 M_Z^4 (12 s_w^6 -32 s_w^4 +29 s_w^2 - 9)
\nn\\
&&
+ 2 M_Z^2 s (s_w^2 - 2)(12 s_w^4 - 12 s_w^2 +1) + 
s^2 (-4 s_w^4+8s_w^2-5) ] \mathcal D_0(s, M_Z^2, M_W^2,M_W^2) 
\nn\\
&&
+ \frac{\alpha \, M_Z^2}{2\,\pi \, s_w^2 \, c_w^2 \, s \, (s-4 M_Z)^2}
\bigg[-4 M_Z^6(s_w^2-1)(4 s_w^2-3)(12 s_w^4 - 20s_w^2 + 9) 
\nn \\
&&
 + 2 M_Z^4 s (18 s_w^4-34s_w^2 + 15)(4(s_w^2-3)s_w^2+7) - 2M_Z^2 s^2 (12 s_w^8-96s_w^6 +201s_w^4-157s_w^2+41)+ 
\nn \\
&&
 s^3(-12 s_w^6+32s_w^4-27s_w^2+7)  \bigg]\, \mathcal C_0(s, M_Z^2,M_Z^2 ,M_W^2,M_W^2,M_W^2)
\eeqa
\beqa
&&
\Xi^{W}_{ZZ}(p,q)
=\frac{\alpha\,M_Z^2}{2 \,\pi\, s_w^2\,c_w^2\,s (s-4M_Z^2)}
\bigg[2 M_Z^4 \left(-12 s_w^6+32 s_w^4-29 s_w^2+9\right)
\nn\\
&&
+ s M_Z^2 \left[4\left(s_w^4+ s_w^2\right)-7\right]- 2 s^2 \left(s_w^2-1\right) \bigg]
+ \frac{\alpha M_Z^2}{\pi \,s_w^2c_w^2 }(-2s_w^4 + 3s_w^2 - 1)\, \mathcal B_0(s,M_W^2,M_W^2)
\nn \\
&&
+ \frac{\alpha \, M_Z^2}{\pi\, s_w^2 \, c_w^2\, s (s-M_Z^2)^2} 
\big[\vphantom{\left(s_w^2-1\right)^2}2 M_Z^6 \left(12 s_w^6-32 s_w^4+29 s_w^2-9\right)
+ s M_Z^4 \left(-24 s_w^6 + 92 s_w^4 - 110 s_w^2 + 41 \right)
\nn\\
&&
+ s^2 M_Z^2 \left(-12 s_w^4+26 s_w^2-13\right)+ 2 s^3 \left(s_w^2-1\right)^2\big]\mathcal D_0(s, M_Z^2, M_W^2,M_W^2)
\nn\\
&&
+ \frac{\alpha\,M_Z^2}{4 \,\pi \,s_w^2\,c_w^2\,s(s-M_Z^2)^2} 
\big[ -8 M_Z^8 \left(s_w^2 - 1\right) \left(4 s_w^2-3\right) \left(12 s_w^4-20 s_w^2+9\right)
\nn \\
&&
+ 4 s M_Z^6 \left(24 s_w^8-60 s_w^6+30 s_w^4+25 s_w^2 - 18 \right)
+ 2 s^2 M_Z^4 (-20 s_w^6+76 s_w^4  - 103 s_w^2+46) 
\nn \\
&&
+ s^3 M_Z^2 \left(-4 s_w^4+24 s_w^2-19\right) - 2 s^4 \left(s_w^2-1\right)\big]\, 
\mathcal C_0(s, M_Z^2,M_Z^2 ,M_W^2,M_W^2,M_W^2)\vphantom{\frac{M_Z^2}{(s-4M_Z^2)}}
\eeqa
%
\item{\bf The $(Z,H)$ sector}
%
\beqa
&&
\Phi^{ZH}_{ZZ}(p,q) 
= 
\frac{- \alpha}{4\,\pi\,s\,c_w^2 s_w^2\, \left( s - 4 M_Z^2 \right)}
\left\{\vphantom{\frac{M_H^2}{\left(s-4 M_Z^2\right)}} \left[M_H^2 \left(s-2 M_Z^2\right)+3 s M_Z^2- 2 M_Z^4\right] 
\right.
\nn\\
&&
\left.
+ 2 \left(M_Z^2-M_H^2\right)\left(\mathcal A_0(M_Z^2)-\mathcal A_0(M_H^2)\right)
\right.
\nn\\
&&
\left.
+ \frac{1}{\left(s-4 M_Z^2\right)} \left[2 M_H^2 \left(s M_Z^2-2 M_Z^4+s^2\right)+3 s^2 M_Z^2-14 s M_Z^4+8 M_Z^6\right]
\mathcal B_0(s, M_Z^2,M_Z^2)
\right.
\nn\\
&&\left. 
- \frac{1}{\left(s-4 M_Z^2\right)}
\left(2 M_H^2+s\right) \left[2 M_H^2 \left(s-M_Z^2\right)-3 s M_Z^2\right]
\mathcal B_0(s, M_H^2,M_H^2)
\right.
\nn\\
&&\left.
+ \frac{2}{\left(s-4 M_Z^2\right)} \left[s M_H^4+6 \left(s-M_H^2\right) M_Z^4+\left(2 M_H^4-3 s M_H^2-3 s^2\right) 
M_Z^2\right]\mathcal B_0(M_Z^2, M_Z^2,M_H^2)
\right.
\nn\\
&&
\left. 
+ \frac{\left(2 M_H^2+s\right)}{\left(s-4 M_Z^2\right)}  \bigg[M_Z^2 \left(-8 s M_H^2-2 M_H^4+s^2\right) 
+ 2 M_Z^4 \left(4 M_H^2 + s\right) + 2 s M_H^4\bigg]\mathcal C_0(s, M_Z^2,M_Z^2 ,M_Z^2,M_H^2,M_H^2) 
\right.
\nn\\
&&
\left.
+ \frac{M_H^2}{\left(s-4 M_Z^2\right)} 
\left[2 M_H^2 \left(s M_Z^2-2 M_Z^4+s^2\right) - 20 s M_Z^4+16 M_Z^6+s^3\right]
\mathcal C_0(s, M_Z^2,M_Z^2 ,M_H^2,M_Z^2,M_Z^2) \right\}  
\eeqa
\beqa
&&
\Xi^{ZH}_{ZZ}(p,q)
= -  \frac{\alpha}{8\, \pi \, s\,c_w^2s_w^2 \left(s-4 M_Z^2\right)}
\bigg\{- 4 M_Z^2 \left(M_Z^4+M_H^2 M_Z^2-3 s M_Z^2+s^2\right)
\nn\\
&&
+ 2 \left(M_H^2-M_Z^2 \right)\left(s-2 M_Z^2\right)\left(\mathcal A_0(M_Z^2)-\mathcal A_0(M_H^2)\right)
\nn\\
&& 
- \frac{1}{\left(s-4 M_Z^2\right)} \left[\left(8 M_Z^6+s^3\right) M_H^2+M_Z^2 \left(s-4 M_Z^2\right) \left(s-2 
M_Z^2\right)\left(3 s-2 M_Z^2\right)\right]\mathcal B_0(s, M_Z^2,M_Z^2)  \nn\\
&&
+ \frac{1}{\left(s-4 M_Z^2\right)} \left[2 \left(4 M_H^4-s^2\right) M_Z^4 
- s \left(2 M_H^2+s \right){}^2 M_Z^2+s^2 M_H^2\left(2 M_H^2+s \right)\right]\, \mathcal B_0(s, M_H^2,M_H^2)
\nn\\
&&
+ \frac{8 M_Z^2}{\left(s-4 M_Z^2\right)} \left[-s M_H^4-\left(3 M_H^2+5 s \right) M_Z^4
+ \left(M_H^2+s \right)\left(M_H^2+2 s \right) M_Z^2\right]\mathcal B_0(M_Z^2, M_Z^2,M_H^2)
\nn\\
&& 
- \frac{\left(2 M_H^2+s \right)}{\left(s-4 M_Z^2\right)} 
\left[4 \left(7 s-4 M_H^2\right) M_Z^6+4 \left(M_H^2-s \right) \left(M_H^2+3 s \right) M_Z^4
\right.
\nn\\
&&
\left. 
+ 2 s \left(-M_H^4-2 s M_H^2+s^2\right) M_Z^2+s^2 M_H^4\right]\mathcal C_0(s, M_Z^2,M_Z^2 ,M_Z^2,M_H^2,M_H^2)
\nn \\
&&
- \frac{1}{\left(s-4 M_Z^2\right)} \left[\left(8 M_Z^6+s^3\right) M_H^4 \right.
\nn\\
&&
\left. 
+ 4 M_Z^2 \left(s-4 M_Z^2\right) \left(2 M_Z^4-s M_Z^2+s^2\right) M_H^2 
\right.
\nn \\
&&
\left.
+ 4 s M_Z^4 \left(s-4 M_Z^2\right){}^2\right]\mathcal C_0(s, M_Z^2,M_Z^2 ,M_H^2,M_Z^2,M_Z^2)
\bigg\}  
\eeqa

\item{\bf Term of improvement}
\beqa
&&
\Phi^{I}_{ZZ}(p,q)
= \frac{3\,\chi\,\alpha}{2\, \pi\, s_w^2\, c_w^2\,(s-4M_Z^2)^2}
\bigg\{ (c_w^2 - s_w^2)^2 \bigg[ 2 M_Z^2 s - 8 M_Z^4 
\nn \\
&&
+ 2 M_Z^2 (s + 2 M_Z^2) \, \mathcal D_0 \left( s, M_Z^2 , M_W^2, M_W^2 \right) 
+ 2 \left( c_w^2 M_Z^2 (8 M_Z^4 - 6 M_Z^2 s + s^2) - 2 M_Z^6 + 2 M_Z^4 s \right)
\nn \\
&& 
\times \, \mathcal C_0 \left( s, M_Z^2, M_Z^2, M_W^2,M_W^2,M_W^2 \right) \bigg]  
+  2 M_Z^2 s - 8 M_Z^4  + 2 M_Z^2 (s + 2\, M_Z^2) \big[
\mathcal B_0 \left(s, M_Z^2, M_Z^2 \right)  \nn \\
&& 
-  \mathcal B_0\left(M_Z^2, M_Z^2, M_H^2 \right)  \big] 
+ \left( 3 M_Z^2 s - 2 M_H^2 (s - M_Z^2) \right) \big[ \mathcal B_0 
\left(s, M_H^2, M_H^2 \right) -  \mathcal B_0 \left(s, M_Z^2, M_Z^2 \right)\big] 
\nn \\
&& 
+ M_H^2 \left( 2 M_H^2 (s-M_Z^2) + 8 M_Z^4 - 6 M_Z^2 s + s^2 \right) \mathcal C_0\left( s, M_Z^2,M_Z^2, M_H^2,M_Z^2,M_Z^2\right) 
\nn \\
&& + \left( 2 M_H^2 (M_H^2 - 4 M_Z^2)(s-M_Z^2) + s M_Z^2 (s+ 2 M_Z^2)\right) \mathcal C_0\left( s, M_Z^2,M_Z^2, 
M_Z^2,M_H^2,M_H^2\right) \bigg\} \\
&&
\Xi^{I}_{ZZ}(p,q)
= - \frac{3\,\chi\,\alpha\,s}{8\,\pi\, s_w^2 \, c_w^2 (s-4 M_Z^2)^2} 
\, \bigg\{ (c_w^2 -s_w^2)^2 \bigg[ 4 M_Z^4 (s - 4 M_Z^2)
\nn \\ 
&&
+ 8 M_Z^4 (s-M_Z^2) \, \mathcal D_0 \left( s, M_Z^2, M_W^2, M_W^2\right)
\nn \\
&&  
+ 2 M_Z^4 \left[ s^2 - 2 M_Z^2 s + 4 M_Z^4 + 4 c_w^2 M_Z^2 (s - 4 M_Z^2) \right]\,\mathcal C_0 \left( s, M_Z^2, M_Z^2, 
M_W^2,M_W^2,M_W^2 \right) \bigg] 
\nn \\
&&  
+ 4 M_Z^2 s_w^4 c_w^2 s (s - 4 M_Z^2)^2 \mathcal C_0 \left( s, M_Z^2, M_Z^2, M_W^2,M_W^2,M_W^2 \right) + 4 M_Z^2 (s-4 M_Z^2)
\nn \\
&&
+ \left[M_Z^2 s (s + 2 M_Z^2) - M_H^2 (s^2 - 2 M_Z^2 s + 4 M_Z^4)\right] 
\big[ \mathcal B_0 \left(s, M_H^2, M_H^2 \right) -  \mathcal B_0 \left(s, M_Z^2, M_Z^2 \right) \big] 
\nn \\
&& 
+ 8 M_Z^4 (s - M_Z^2) \big[ \mathcal B_0 \left(s, M_Z^2, M_Z^2 \right) -  \mathcal B_0 \left(M_Z^2, M_Z^2, M_H^2 \right)  \big] 
+ M_H^2 \left[ 4 M_Z^4 (s-4 M_Z^2)
\right. 
\nn \\
&&
\left. 
+ M_H^2 (s^2 - 2 MZ^2 s + 4 M_Z^4)\right] \mathcal C_0\left( s, M_Z^2,M_Z^2, M_H^2,M_Z^2,M_Z^2\right) + \left[ M_H^2(M_H^2 - 
4 M_Z^2) (s^2 - 2 M_Z^2 s + 4 M_Z^4) 
\right.
\nn \\
&&
\left. 
+ 2 M_Z^2 s (s^2 - 6 M_Z^2 s + 14 M_Z^4)\right] \mathcal C_0\left( s, M_Z^2,M_Z^2, M_Z^2,M_H^2,M_H^2 \right)
\bigg\}.
\eeqa

\item{\bf External leg corrections}
\end{itemize}
%
The $\Delta^{\alpha\beta}(p,q)$ correlator is decomposed as
\beqa
\Delta^{\alpha\beta}(p,q) 
&=& 
\bigg[ \Sigma_{Min, \, \rho H}(k^2) + \Sigma_{I,\,\rho H}(k^2) \bigg] 
\frac{1}{s - M_H^2} V_{HZZ}^{\alpha\beta} + \left( \frac{\Lambda}{i} \right) V_{I, \, \rho H}(k) \frac{1}{s - M_H^2} 
\Sigma_{HH}(k^2) \frac{1}{s - M_H^2} 
V_{HZZ}^{\alpha\beta}  
\nn \\
&+& 
\Delta^{\alpha\beta}_{I, \, HZZ}(p,q)
\eeqa
where $\Sigma_{HH}(k^2)$ is the Higgs self-energy, $V_{HZZ}^{\alpha\beta}$ and
$ V_{I, \, \rho H}$ are tree level vertices defined in appendix (\ref{rules}) and $\Delta^{\alpha\beta}_{I, \, HZZ}(p,q)$ 
is expanded into the three contributions of improvement as
\beqa
\Delta^{\alpha\beta}_{I, \, HZZ}
&=&  
\Delta^{\alpha\beta}_{(F), \, HZZ}(p,q) + \Delta^{\alpha\beta}_{(W), \, HZZ}(p,q) + \Delta^{\alpha\beta}_{(Z,H), \, HZZ}(p,q)
\nonumber \\
&=&
\left[ \left( \frac{s}{2} - M_Z^2 \right) \eta^{\alpha\beta} - q^\alpha p^\beta \right]\, \Phi^{\Delta}_{ZZ}(p,q) +
\eta^{\alpha\beta}\, \Xi^{\Delta}_{ZZ}(p,q)
\nn \\
&=& 
\left[ \left( \frac{s}{2} - M_Z^2 \right) \eta^{\alpha\beta} - q^\alpha p^\beta \right]
\left(\Phi^{\Delta\,F}_{ZZ}(p,q) + \Phi^{\Delta\,W}_{ZZ}(p,q) + \Phi^{\Delta\,W}_{ZZ}(p,q)\right)  
\nn \\
&& \hspace{35mm}
+\, \eta^{\alpha\beta}\, \left(\Xi^{\Delta\,F}_{ZZ}(p,q) + \Xi^{\Delta\,W}_{ZZ}(p,q) + \Xi^{\Delta\,W}_{ZZ}(p,q) \right)\, .
\eeqa
These are given by
\bea
&&
\Phi^{\Delta\,F}_{ZZ}(p,q) = 
- \sum_f \frac{6 \, \alpha \,\chi \, m_f^2}{\pi\, s_w^2 \, c_w^2 \, (s-M_H^2)(s - 4 M_Z^2)}
\bigg\{
( C_v^{f\, 2} + C_a^{f\, 2}) (s-2 M_Z^2)  \nn \\
&& 
+  \frac{2}{s-4 M_Z^2} [ M_Z^2 ( C_v^{f\, 2} + C_a^{f\, 2}) (s +2 M_Z^2) + C_a^{f \, 2} (s-4M_Z^2) s ] \mathcal 
D_0(s,M_Z^2,m_f^2,m_f^2) \nn \\
&&
+ \frac{s - 2 M_Z^2}{2(s-4 M_Z^2)} [ ( C_v^{f\, 2} + C_a^{f\, 2}) (4 m_f^2 (s-4 M_Z^2)  + 4 M_Z^4 + 6 M_Z^2 s - s^2 ) + 2 C_a^{f \, 
2} s (s-4 M_Z^2) ] \times \nn \\
&&
\mathcal C_0(s, M_Z^2,M_Z^2,m_f^2,m_f^2,m_f^2) 
\bigg\} \, , \\
&&
\Xi^{\Delta\,F}_{ZZ}(p,q)
= \sum_f \frac{18\,\alpha\,s\,\chi \, m_f^2}{\pi \,  s_w^2 \, c_w^2 \, (s - M_H^2)(s -4 M_Z^2) } \, 
\eta^{\alpha \beta} 
\bigg\{
2 M_Z^2 ( C_v^{f\, 2} + C_a^{f\, 2}) \nn \\
&&
+ 2 s (s- 4 M_Z^2) C_a^{f \, 2} \mathcal B_0(s, m_f^2, m_f^2) + \frac{2}{s- 4M_Z^2} [ 2 M_Z^4 ( C_v^{f\, 2} + C_a^{f\, 2}) (s - 
M_Z^2) + C_a^{f \, 2} M_Z^2 (s-4M_Z^2) s ] \times \nn \\
&&
\mathcal D_0(s,M_Z^2,m_f^2,m_f^2)
+
\frac{1}{s -4 M_Z^2} [  ( C_v^{f\, 2} + C_a^{f\, 2}) M_Z^4 ( 4 m_f^2 (s -4 M_Z^2) + 4 M_Z^4 - 2 M_Z^2 s + s^2)  \nn \\
&&
+ 2 C_a^{f \, 2} s (s - 4 M_Z^2) (m_f^2 (s- 4 M_Z^2) + M_Z^4)] \mathcal C_0(s, M_Z^2,M_Z^2,m_f^2,m_f^2,m_f^2) 
\bigg\}
\eea
\bea
&&
\Phi^{\Delta\,W}_{ZZ}(p,q) = 
\frac{3\,\alpha \, \chi }{\pi\, s_w^2 \, c_w^2 \, (s-M_H^2)(s-4 M_Z^2)} 
\bigg\{
\frac{s - 2 M_Z^2}{2} [ M_H^2 (1- 2 s_w^2)^2 \nn \\
&&
+ 2 M_Z^2 (-12 s_w^6 + 32 s_w^4 - 29 s_w^2 + 9)]
+ \frac{M_Z^2}{s - 4 M_Z^2} [ M_H^2 (1 - 2 s_w^2)^2 (s + 2 M_Z^2) \nn \\
&&
- 2 (s_w^2 -1) ( 2 M_Z^4 (12 s_w^4 - 20 s_w^2 + 9) + s M_Z^2 (12 s_w^4 - 20 s_w^2 +1 ) + 2 s^2) ] \mathcal D_0(s, M_Z^2, M_W^2,M_W^2) 
\nn \\
&&
+ \frac{M_Z^2}{s - 4 M_Z^2} [ 2 (s_w^2 -1) ( 2 M_Z^6 (4 s_w^2-3)(12 s_w^4 - 20 s_w^2 + 9) 
+ 2 M_Z^4 s (- 36 s_w^6 + 148 s_w^4 - 163 s_w^2 + 54) \nn \\
&&
+ M_Z^2 s^2 (12 s_w^6 - 96 s_w^4 + 125 s_w^2 - 43) + 4 s^3 ( 2 s_w^4 - 3 s_w^2 + 1) ) - M_H^2 (1- 2 s_w^2)^2 (M_Z^4 (8 s_w^2 -6) \nn 
\\
&&
+ 2 M_Z^2 s (2 - 3 s_w^2) + s^2 (s_w^2-1))  ] \mathcal C_0(s,M_Z^2,M_Z^2,M_W^2,M_W^2,M_W^2) 
\bigg\}\, ,  \\
&&
\Xi^{\Delta\,W}_{ZZ}(p,q)
=\frac{3\,\alpha\, \chi \, M_Z^2}{\pi\,s_w^2 \, c_w^2 \, (s-M_H^2)(s-4 M_Z^2)}
\bigg\{
- M_Z^2 [ M_H^2 (1- 2 s_w^2)^2 + 2 M_Z^2 (-12 s_w^6 + 32 s_w^4 - 29 s_w^2 + 9)] \nn \\
&&
+ \frac{s (s- 4 M_Z^2)}{2} (8 s_w^4 - 13 s_w^2 + 5) \mathcal B_0(s, M_W^2,M_W^2) 
+ \frac{2}{s - 4 M_Z^2} [ M_H^2 M_Z^2 (1-2s_w^2)^2 (M_Z^2- s ) \nn \\
&&
- 2 (s_w^2 -1)(M_Z^6 (12 s_w^4 - 20 s_w^2 + 9)  - 3 M_Z^4 s ( 4 (s_w^2 -3)s_w^2 + 7) + M_Z^2 s^2 (7 - 8 s_w^2) + s^3 (s_w^2 -1)  )] 
\times \nn \\ 
&&
\mathcal D_0(s, M_Z^2, M_W^2,M_W^2)
+ \frac{1}{2 (s-4 M_Z^2)} [  M_H^2 (- 4 M_Z^6 (1 -2 s_w^2 )^2 (4 s_w^2-3) + 2 M_Z^4 s (24 s_w^6 - 28 s_w^4 + 6 s_w^2 - 1)  \nn \\
&&
+ M_Z^2 s^2 (- 16 s^6 + 12 s_w^4 + 4 s_w^2 -1) + 2 s^3 s_w^4 (s_w^2 -1)  )  + 2 (s_w^2 -1) ( 4 M_Z^8 ( 4 s_w^2 -3)(12 s_w^4 - 20 
s_w^2 + 9) \nn \\
&&
- 2 M_Z^6 s (24 s_w^6 - 52 s_w^4 + 6 s_w^2 + 15) + M_Z^4 s^2 (45 - 4 s_w^2 (s_w^2+ 13)) + 2 M_Z^2 s^3 (4 s_w^4 + 2 s_w^2 -5) \nn \\
&&
- s^4(s_w^4 -1) ) ]  \mathcal C_0(s,M_Z^2,M_Z^2,M_W^2,M_W^2,M_W^2) 
\bigg\}
\\
&&
\Phi^{\Delta\,ZH}_{ZZ}(p,q) =  
\frac{\alpha \, \chi}{\pi\,s_w^2\,c_w^2\,(s - M_H^2)(s- 4 M_Z^2)}\,
\bigg\{
(2 M_H^2 + M_Z^2) (s - 2 M_Z^2) \nn \\
&&
- 2 (M_H^2 - M_Z^2) (\mathcal A_0(M_Z^2) - \mathcal A_0(M_H^2)) + \frac{1}{s - 4 M_Z^2} [2 M_H^4 (s - M_Z^2) + 3 M_H^2 M_Z^2 s \nn \\
&&
+ 2 M_Z^2 (4 M_Z^4 - 9 M_Z^2 s + 2 s^2)] \mathcal B_0(s, M_Z^2, M_Z^2) - \frac{3}{s - 4 M_Z^2} [2 M_H^4(s - M_Z^2)- 3 M_H^2 M_Z^2 s] 
\mathcal B_0(s, M_H^2,M_H^2)  \nn \\
&&
- \frac{2}{s-4 M_Z^2} [ M_H^2 (s + 2 M_Z^2)(4 M_Z^2 - M_H^2) + 2 M_Z^2 s (s - 4 M_Z^2)] \mathcal B_0(M_Z^2, M_Z^2, M_H^2) \nn \\
&&
- \frac{3 M_H^2}{s- 4 M_Z^2} [2 M_H^2 (s-M_Z^2)(4 M_Z^2 - M_H^2) - M_Z^2 s (s + 2 M_Z^2)] \mathcal 
C_0(s,M_Z^2,M_Z^2,M_Z^2,M_H^2,M_H^2)  + \frac{M_H^2}{s - 4 M_Z^2} \times \nn \\
&&
[ 2 M_H^4 (s-M_Z^2) + M_H^2 (4 M_Z^4 - 2 M_Z^2 s + s^2) + 2 M_Z^2 (8 M_Z^4 - 14 M_Z^2 s + 3 s^2)] \mathcal 
C_0(s,M_Z^2,M_Z^2,M_H^2,M_Z^2,M_Z^2)
\bigg\} 
\nn \\
\eeqa
\beqa
&&
\Xi^{\Delta\,ZH}_{ZZ}(p,q)
= - \frac{3\,\alpha\,\chi}{\pi\,s_w^2\,c_w^2\,(s- M_H^2)(s -4 M_Z^2)}\,
\bigg\{ M_Z^4 (2 M_H^2 + M_Z^2) 
\nn \\
&&
+ \frac{1}{2} (M_Z^2 - M_H^2) (s - 2 M_Z^2)  (\mathcal A_0(M_Z^2) - \mathcal A_0(M_H^2))
+ \frac{1}{4 (s - 4 M_Z^2)}\, [ M_H^4 (4 M_Z^4 - 2 M_Z^2 s + s^2)
\nn \\
&& 
+ M_H^2 M_Z^2 s (s + 2 M_Z^2) - M_Z^2 (16 M_Z^6 - 28 M_Z^4 s + 18 
M_Z^2 s^2 - 3 s^3)] \, \mathcal B_0(s, M_Z^2, M_Z^2)
\nn \\
&&
- \frac{3 M_H^2}{4(s-4 M_Z^2)} [M_H^2 (4 M_Z^4 - 2 M_Z^2 s + s^2) - M_Z^2 s (s + 2 M_Z^2)] \mathcal B_0 
(s, M_H^2, M_H^2) \nn \\
&&
+ \frac{2}{s- 4 M_Z^2} [ M_Z^2 s (M_H^2 - 2 M_Z^2)^2 - M_H^2 M_Z^4 (M_H^2 - 4 M_Z^2) - M_Z^4 s^2] \mathcal B_0(M_Z^2, M_Z^2, M_H^2)
\nn \\
 &&
+ \frac{3 M_H^2}{4 (s- 4 M_Z^2)} \times
[ M_H^2 (4 M_Z^4 - 2 M_Z^2 s + s^2) (M_H^2 - 4 M_Z^2) + 2 M_Z^2 s (16 M_Z^4 - 6 M_Z^2 s + s^2)]\,\times 
\nn \\ 
&&
\mathcal C_0(s, M_Z^2, M_Z^2, M_Z^2, M_H^2, M_H^2)
+ \frac{1}{4(s - 4 M_Z^2)} [ M_H^6 (4 M_Z^4 - 2 M_Z^2 s + s^2) + 2 M_H^4 M_Z^2 (s^2 - 4 M_Z^4) \nn \\
&&
- 4 M_H^2 M_Z^2 (8 M_Z^6 - 10 M_Z^4 s + 6 M_Z^2 s^2 - s^3)
+ 4 M_Z^4 s (s - 4 M_Z^2)^2 ] \mathcal C_0(s, M_Z^2, M_Z^2, M_H^2, M_Z^2, M_Z^2)
\bigg\}
\eea
\bea
\Sigma_{Min, \, \rho H}(k^2) 
&=& 
\frac{e }{48 \pi^2 \, s_w \, c_w \, M_Z}  \bigg\{ \sum_f m_f^2 \bigg[ 3 (4 m_f^2 - k^2) \mathcal 
B_0(k^2,m_f^2,m_f^2) + 12 \mathcal A_0(m_f^2) -2 k^2 + 12 m_f^2 \bigg] \nn \\
&-& 
\frac{1}{2} \bigg[ 3 \left( k^2 (M_H^2 -6 M_W^2)  + 2 M_W^2 (M_H^2 + 6 M_W^2) \right) \mathcal B_0(k^2, M_W^2,M_W^2) + 6 (M_H^2 + 6 
M_W^2) \mathcal A_0(M_W^2) \nn \\
&-& k^2 (M_H^2 + 18 M_W^2) + 6 M_W^2 (M_H^2 - 2 M_W^2) \bigg] 
- \frac{1}{4} \bigg[ 3 \left(M_H^2 (2 M_Z^2 + s) + 12 M_Z^4 - 6 M_Z^2 s \right) \nn \\
&\times& \mathcal B_0(s, M_Z^2,M_Z^2) + 9 M_H^2 (2 M_H^2 + s) \mathcal B_0(s, M_H^2,M_H^2) + 6 (M_H^2 + 6 M_Z^2) \mathcal A_0(M_Z^2) 
+ 18 M_H^2 \mathcal A_0(M_H^2) \nn \\
&+&  
2 (9 M_H^4 + M_H^2 (3 M_Z^2 -2 s) - 6 M_Z^4 - 9 M_Z^2 s ) \bigg]
\bigg\} 
\eea
\bea
\Sigma_{I, \, \rho H}(k^2) &=& \frac{3 e}{16 \pi^2 \, s_w \, c_w} \frac{k^2 \, M_H^2}{M_Z^2} \chi \, \bigg[ \mathcal B_0(k^2, M_W^2, 
M_W^2) + \frac{3}{2} \mathcal B_0(k^2,M_H^2,M_H^2)  + \frac{1}{2} \mathcal B_0(k^2, M_Z^2, M_Z^2) \bigg]
\eea

\section{Appendix. Standard Model self-energies}
\label{SigmaSM}
%
%
We report here the expressions of the self-energies appearing in Section \ref{renorm} which define the renormalization conditions. They are given by
\bea
\Sigma^{\gamma\gamma}_T(p^2)
&=& - \frac{\alpha}{4 \pi} \bigg\{ \frac{2}{3} \sum_f \, N_C^f 2 Q_f^2 \bigg[
-(p^2 + 2 m_f^2)B_0(p^2, m_f^2, m_f^2) Ê+ Ê2 m_f^2 B_0(0, m_f^2, m_f^2) + \frac{1}{3}p^2 \bigg] \nn \\
&+& Ê\bigg[ (3 p^2 + 4 M_W^2 ) B_0(p^2, M_W^2, M_W^2) - 4 M_W^2 B_0(0, M_W^2, M_W^2)\bigg]\bigg\}
\eea
\bea
\Sigma^{ZZ}_T(p^2)
&=& -\frac{\alpha}{4 \pi} \bigg\{ \frac{2}{3} \sum_f \,
N_C^f \bigg[ \frac{C_V^{f \, 2} + C_A^{f \, 2}}{2 s_w^2 c_w^2}
\bigg( -(p^2 + 2m_f^2) B_0(p^2, m_f^2, m_f^2) Ê+ Ê2 m_f^2 B_0(0, m_f^2, m_f^2) + \frac{1}{3}p^2 \bigg) \nn\\
&+& \frac{3}{4 s_w^2 c_w^2} m_f^2 B_0(p^2,m_f^2, m_f^2) \bigg]
 + Ê\frac{1}{6 s_w^2 c_w^2}\bigg[ \bigg( (18 c_w^4 + 2 c_w^2 -\frac{1}{2})p^2 + (24 c_w^4 + 16 c_w^2 -10)M_W^2 \bigg) \nn \\
&\times& B_0(p^2, M_W^2, M_W^2) Ê- Ê(24 c_w^4 - 8 c_w^2 + 2)M_W^2 B_0(0, M_W^2, M_W^2) + (4 c_w^2-1)\frac{p^2}{3} Ê\bigg] \nn \\
&+& Ê \frac{1}{12 s_w^2 c_w^2} \bigg[ (2 M_H^2 -10 M_Z^2 - p^2) B_0(p^2, M_Z^2, M_H^2) Ê- Ê 2 M_Z^2 B_0(0, M_Z^2, M_Z^2) - 2 M_H^2 B_0(0, M_H^2, M_H^2) Ê\nn \\
&& - Ê\frac{(M_Z^2 - M_H^2)^2}{p^2}\left( B_0(p^2, M_Z^2, M_H^2) Ê- B_0(0, M_Z^2, M_H^2) \right) -
\frac{2}{3} p^2 Ê \bigg] \bigg\}
\eea
\bea
\Sigma^{\gamma Z}_T(p^2)
&=& \frac{\alpha}{4 \pi \, s_w \, c_w} \bigg\{ \frac{2}{3} \sum_f \, N_C^f \,
Q_f \, C_V^f \bigg[ (p^2 + 2m_f^2) B_0(p^2, m_f^2, m_f^2) - 2 m_f^2 B_0(0, m_f^2, m_f^2) -\frac{1}{3}p^2 \bigg] \nn \\
&-& \frac{1}{3} \bigg[ \left( (9 c_w^2 + \frac{1}{2})p^2
+ (12 c_w^2 + 4)M_W^2 \right) B_0(p^2, M_W^2, M_W^2) - Ê(12 c_w^2 -2)M_W^2 B_0(0, M_W^2, M_W^2) + \frac{1}{3}p^2 \bigg]\bigg\} \nn \\
\eea
\bea
\Sigma_{HH}(p^2) &=& - \frac{\alpha}{4 \pi} \bigg\{ \sum_{f} N_C^f \frac{m_f^2}{2 s_w^2 M_W^2} \bigg[ 2 \mathcal A_0\left( m_f^2 \right) + (4 m_f^2 - p^2) \mathcal B_0 \left( p^2, m_f^2,m_f^2\right) \bigg] \nn \\
&-& \frac{1}{2 s_w^2} \bigg[ \left(6 M_W^2 - 2p^2 + \frac{M_H^4}{2 M_W^2} \right) \mathcal B_0 \left( p^2, M_W^2, M_W^2 \right) + \left( 3 + \frac{M_H^2}{2 M_W^2} \right) \mathcal A_0 \left( M_W^2 \right) - 6 M_W^2 \bigg] \nn \\
&-& \frac{1}{4 s_w^2 \, c_w^2} \bigg[ \left(6 M_Z^2 - 2p^2 + \frac{M_H^4}{2 M_Z^2} \right) \mathcal B_0 \left( p^2, M_Z^2, M_Z^2 \right) + \left( 3 + \frac{M_H^2}{2 M_Z^2} \right) \mathcal A_0 \left( M_Z^2 \right) - 6 M_Z^2 \bigg] \nn \\
&-& \frac{3}{8 s_w^2} \bigg[ 3 \frac{M_H^4}{M_W^2} \mathcal B_0 \left( p^2, M_H^2,M_H^2 \right) + \frac{M_H^2}{M_W^2} \mathcal A_0 \left( M_H^2 \right) \bigg] \bigg\}
\eea
\bea
\Sigma^{WW}_T(p^2) &=& -\frac{\alpha}{4 \pi} \bigg\{ \frac{1}{3 s_w^2} \sum_i \bigg[ \left( \frac{m_{l,i}^2}{2} - p^2 \right) 
\mathcal B_0\left(p^2, 0, m_{l,i}^2 \right) + \frac{p^2}{3} + m_{l,i}^2 \mathcal B_0 \left( 0, m_{l,i}^2,m_{l,i}^2 \right) Ê\nn \\
&+& 
Ê\frac{m_{l,i}^4}{2 p^2} \left( \mathcal B_0\left(p^2, 0, m_{l,i}^2 \right) - \mathcal B_0\left(0, 0, m_{l,i}^2 \right)\right) \bigg] 
+ \frac{1}{s_w^2} \sum_{i,j}|V_{ij}|^2 \bigg[ 
\left( \frac{m_{u,i}^2 + m_{d,j}^2}{2} - p^2\right) \mathcal B_0 \left( p^2, m_{u,i}^2, m_{d,j}^2\right) \nn \\
&+& 
\frac{p^2}{3} + m_{u,i}^2 \mathcal B_0 \left( 0, m_{u,i}^2, m_{u,i}^2\right) + m_{d,j}^2 \mathcal B_0 \left( 0, m_{d,j}^2, 
m_{d,j}^2\right) + \frac{(m_{u,i}^2 - m_{d,j}^2)^2}{2 p^2} \big( \mathcal B_0 \left( p^2, m_{u,i}^2, m_{d,j}^2\right) \nn \\
&-& 
\mathcal B_0 \left( 0, m_{u,i}^2, m_{d,j}^2\right)\big) \bigg] Ê + \frac{2}{3} \bigg[ (2 M_W^2 + 5 p^2) \mathcal B_0 \left( p^2, 
M_W^2, \lambda^2 \right) - 2 M_W^2 \mathcal B_0 \left( 0, M_W^2, M_W^2\right) \nn \\
&-&
\frac{M_W^4}{p^2} \big( \mathcal B_0\left( p^2, M_W^2, \lambda^2 \right) - \mathcal B_0 \left( 0,M_W^2, \lambda^2 \right) \big) + 
\frac{p^2}{3} Ê\bigg] + \frac{1}{12 s_w^2} \bigg[
\big( (40 c_w^2 -1)p^2 \nn \\
&+& 
(16 c_w^2 + 54 - 10 c_w^{-2}) M_W^2 \big) \mathcal B_0 \left(p^2, M_W^2, M_Z^2 \right) - (16 c_w^2 + 2) \big( M_W^2 \mathcal B_0 
\left( 0,M_W^2,M_W^2\right) \nn \\
&+& 
M_Z^2 \mathcal B_0 \left( 0, M_Z^2, M_Z^2\right) \big) + (4 c_w^2 -1) \frac{2 p^2}{3} - Ê(8 c_w^2 +1) \frac{(M_W^2 - 
M_Z^2)^2}{p^2} \big( \mathcal B_0 \left( p^2, M_W^2,M_Z^2\right) \nn \\
&-& 
\mathcal B_0 \left(0, M_W^2,M_Z^2 \right) \big) \bigg] + \frac{1}{12 s_w^2} \bigg[ (2 M_H^2 - 10 M_W^2 - p^2) \mathcal B_0 
\left(p^2, M_W^2,M_H^2 \right) - 2 M_W^2 \mathcal B_0 \left(0,M_W^2,M_W^2 \right) \nn \\
&-& 
2 M_H^2 \mathcal B_0 \left( 0, M_H^2,M_H^2\right) - \frac{(M_W^2 -M_H^2)^2}{p^2} \big( \mathcal B_0 \left( p^2, M_W^2, 
M_H^2\right) - \mathcal B_0 \left( 0,M_W^2,M_H^2\right) \big) - \frac{2 p^2}{3}\bigg] \bigg\}\, ,
\eea
where the subscripts $l$, $u$ and $d$ stand for "leptons", "up" and "down" (quarks) respectively.
The sum runs over the three generations and $\lambda$ is the photon mass introduced to regularize the infrared
divergencies.


\begin{thebibliography}{10}

\bibitem{Han:1998sg}
T.~Han, J.~D. Lykken, and R.-J. Zhang,
\newblock Phys.Rev. {\bf D59}, 105006 (1999), arXiv:hep-ph/9811350.

\bibitem{Giudice:2000av}
G.~F. Giudice, R.~Rattazzi, and J.~D. Wells,
\newblock Nucl.Phys. {\bf B595}, 250 (2001), arXiv:hep-ph/0002178.

\bibitem{Coriano:2011zk}
C.~Corian\`{o}, L.~Delle~Rose, and M.~Serino,
\newblock Phys.Rev. {\bf D83}, 125028 (2011), arXiv:1102.4558.

\bibitem{Giannotti:2008cv}
M.~Giannotti and E.~Mottola,
\newblock Phys. Rev. {\bf D79}, 045014 (2009), arXiv:0812.0351.

\bibitem{Armillis:2009pq}
R.~Armillis, C.~Corian\`{o}, and L.~Delle~Rose,
\newblock Phys. Rev. {\bf D81}, 085001 (2010), arXiv:0910.3381.

\bibitem{Knecht:2003xy}
M.~Knecht, S.~Peris, M.~Perrottet, and E.~de~Rafael,
\newblock JHEP {\bf 03}, 035 (2004), arXiv:hep-ph/0311100.

\bibitem{Jegerlehner:2005fs}
F.~Jegerlehner and O.~V. Tarasov,
\newblock Phys. Lett. {\bf B639}, 299 (2006), arXiv:hep-ph/0510308.

\bibitem{Armillis:2009sm}
R.~Armillis, C.~Corian\`o, L.~Delle~Rose, and M.~Guzzi,
\newblock JHEP {\bf 12}, 029 (2009), arXiv:0905.0865.

\bibitem{Horejsi:1997yn}
J.~Horejsi and M.~Schnabl,
\newblock Z. Phys. {\bf C76}, 561 (1997), arXiv:hep-ph/9701397.

\bibitem{Armillis:2010qk}
R.~Armillis, C.~Corian\`o, and L.~Delle~Rose,
\newblock Phys.Rev. {\bf D82}, 064023 (2010), arXiv:1005.4173.

\bibitem{Goldberger:2007zk}
W.~D. Goldberger, B.~Grinstein, and W.~Skiba,
\newblock Phys.Rev.Lett. {\bf 100}, 111802 (2008), arXiv:0708.1463.

\bibitem{Denner:1991kt}
A.~Denner,
\newblock Fortschr. Phys. {\bf 41}, 307 (1993), arXiv:0709.1075.

\end{thebibliography}
\end{document}